\documentclass[aps,prb,twocolumn,reprint]{revtex4-1}
\usepackage{graphics}
\usepackage{epstopdf}
\usepackage{epsfig}
\usepackage{graphicx}
\usepackage{epsf,epic}
\usepackage{color}
\usepackage{subfig}
\usepackage{marvosym }
\usepackage{amsmath}
\usepackage{booktabs}
\usepackage{multirow}
\usepackage{physics}
\usepackage{amssymb}
\usepackage{hyperref}
\usepackage{amsfonts}
\usepackage{wrapfig}
\usepackage{pstricks}
\usepackage{multirow}
\usepackage{fancyref}
\usepackage{pst-node}
\usepackage{bm}
\usepackage{dcolumn}
\newcommand{\etal}{\textit{et al.\ }}
\newcommand{\ie}{\textit{i.e.\ }}

\def\dprime{{\prime\prime}}

\graphicspath{{figs/}}
\makeatletter
\usepackage{etoolbox} 
\appto{\appendix}{%
	\@ifstar{\def\theequation@prefix{A.}}%
	{}%
}
\makeatother
\begin{document}
\title{Topological band structure transitions in honeycomb antimonene as
function of buckling}
\author{\href{https://www.santoshkumarradha.me}{Santosh Kumar Radha }}
\email{Corresponding author:santosh.kumar.radha@case.edu}

\author{Walter R. L. Lambrecht}
\email{walter.lambrecht@case.edu}
\affiliation{Department of Physics, Case Western Reserve University, 10900 Euclid Avenue, Cleveland, OH-44106-7079}
\begin{abstract}
 The electronic band topology of 
 monolayer $\beta$-Sb (antimonene) is studied from the flat honeycomb to the
 equilibrium buckled structure
 using first-principles calculations and  analyzed
 using  a tight-binding model and low energy Hamiltonians.
 In flat monolayer Sb, the Fermi level occurs near the intersection of
 two warped
 Dirac cones, one associated with the $p_z$-orbitals, and one with the
 $\{p_x,p_y\}$-orbitals. The differently oriented threefold warping of these two
 cones leads to an unusually shaped  nodal line, which 
 leads to anisotropic in-plane transport properties and goniopolarity.
 A slight buckling opens a gap along the nodal line except
 at six remaining Dirac points, protected by symmetry. Under increasing
 buckling, pairs of Dirac points of opposite winding
 number annihilate at a critical buckling angle. At a second
 critical angle, the remaining Dirac points disappear
 when the band structure becomes a trivial semiconductor. Spin-orbit
 coupling and edge states are discussed. 
\end{abstract}
\maketitle

Since the discovery of graphene,\cite{Novoselov04} the world of 2D
atomically thin materials keeps expanding.  
Of special interest are the elemental 2D materials, such
as silicene, germanene, and the recently realized and earlier theoretically
predicted antimonene and arsenene.\cite{Ares18review,Tsai2016,Ji2016,Akturk15,Wang2015,Zhang15,Kamal15}
Unlike their isovalent analog
phosphorene (monolayer black phosphorus)\cite{li2014} which has a more complex
buckled structure with fourfold symmetry, monolayer
Sb and As are found to prefer the buckled honeycomb structure,
known as $\beta$-Sb, which is also found in silicene and germanene.\cite{Sone2014,Davila2016}
Interestingly, an almost completely flat honeycomb form was reported to be
stabilized epitaxially on a Ag(111) substrate.\cite{Shao18} 
Thus, flat monolayer Sb and As may be the closest analog to graphene but
with the interesting difference that there is one additional valence
electron which places the Fermi level in between the usual $p_z$
derived Dirac point  at $K$ (as in graphene)
and a higher lying $\{p_x,p_y\}$ derived Dirac point.

The electronic structure studies thus far report an indirect band gap
for equilibrium buckled $\beta$-Sb but which undergoes a transition
to a semimetallic state under tensile in-plane strain.\cite{Chuang13,LuHao16}
It is related to a transition
from a trivial to a non-trivial band gap inversion at $\Gamma$.
Topological aspects of the band structure of various group-IV and V
systems were studied by Huang \etal\cite{Huang14} and were also
studied in few layer Sb films as function of thickness.\cite{Bian12,PengFei12,Zhao2015} Flat honeycomb Sb was  shown by Hsu \etal\cite{Hsu2016} to be
a topological crystalline insulator. 

Unlike most of these previous works, we here start from  the completely
flat honeycomb Sb monolayer and explore systematically how
its band structure and topology change as the buckling angle is
gradually increased.
We will show that the Fermi level position
near the intersection of two Dirac cones leads to a number of
interesting topological features, from a uniquely shaped nodal
line to several new Dirac points which are allowed to move
as buckling increases and can mutually annihilate in pairs beyond
a critical buckling angle.

Our study is
carried out using first-principles density functional theory (DFT)
and quasiparticle many-body perturbation theory (MBPT) calculations,
with details given in the computational methods. 
A nearest-neighbor tight-binding
model is used to analyze the results and effective low energy Hamiltonians
describing the band features of interest are presented.

Although our predictions here are theoretical, we note that the possibility
of stabilizing monolayer Sb in its flat form by epitaxy on Ag(111)
has already been demonstrated.\cite{Shao18} In Supplementary Information (SI)
Section II,
we show using DFT calculations  that the flat honeycomb
form of monolayer Sb and As is indeed
a metastable phase but also show that the band structure features of Sb
can still be readily recognized  when Sb is placed on top of Ag.

\begin{figure*}[t]
  \includegraphics[width=18 cm]{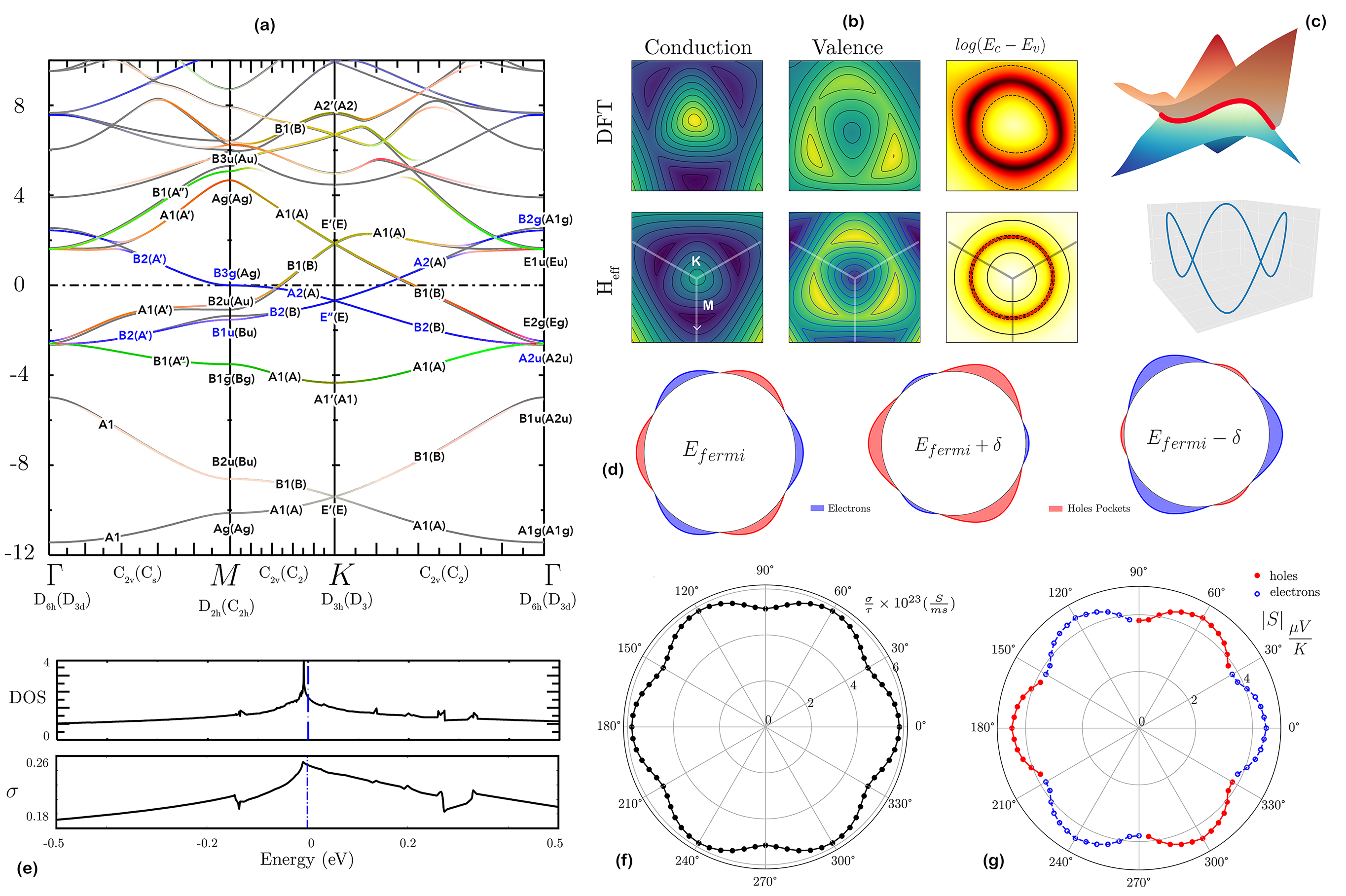}
  \caption{(a) Symmetry labeled LDA band structure of flat (colored) and slightly buckled (grey) monolayer Sb; (b) Contour plots of two Dirac cones and their intersection (on log scale) in DFT (top) and low energy effective hamiltonian (bottom); (c) 3D view of (DFT) intersecting Dirac cones and nodal line; (d) Fermi surface portions around each point $K$, (e) DOS and $\sigma(E)/\tau$ (arb. units)
    (f) Angular dependence of diffuse conductivity
    $\sigma(E_F)/\tau$ and (g) absolute value of thermopower 
    with sign indicated by color (red $<0$, blue $>0$) \label{fig:conesflat}}
\end{figure*}
We start our discussion with the band structure of flat and slightly
buckled monolayer Sb, as shown in Fig. \ref{fig:conesflat}(a) 
and obtained in the generalized gradient approximation (GGA) to DFT.
The symmetry labeling of the bands is crucial to our understanding of
the protection of the  Dirac cones to be discussed.
To make this symmetry labeling unambiguous, it is necessary to
describe the symmetry operations and character tables in detail, which
is done in the  SI-Section III.  The pointgroups of {\bf k}
applying along each symmetry line in the flat and buckled
(labels in parentheses) case are given at the bottom of
Fig. \ref{fig:conesflat}(a). 

Before proceeding with our study of the topological features of interest,
we first point out some similarities and differences of the Sb band structure with the well known band structure of graphene. We immediately recognize
the Dirac point at $K$, here labeled $E^\dprime$ corresponding
to the $p_z$-derived bands (shown in blue) with $z$ perpendicular to the layer.
In graphene, the Fermi level falls at this point but here,
because of the additional valence
electron, it lies higher in energy  shown as the dash-dotted line and
chosen as our reference energy. Another important difference with graphene
is that the $s$-orbitals form a separate set of bands at lower energy
rather than forming strongly hybridized sp$^2$ $\sigma$-bands. This results
from a larger $E_p-E_s$ atomic energy difference relative to the hopping
interactions between the sites. The $\{p_x,p_y\}$ derived bands form 
a separate set of bands (indicated in red ($p_x$), green ($p_y$)
and their mixture)
with another Dirac cone  $E^\prime$ at $K$ above the Fermi level.
The band structure of $\{p_x,p_y\}$ derived bands on the honeycomb lattice
was discussed by Wu and Das Sarma\cite{wu_sarma_2008} in a tight-binding
approximation relevant to optical lattices where only the $\sigma$
interaction is non zero.  Here, both the $V_\sigma$ and $V_\pi$ matter.
While in a tight-binding model
of each set of bands separately, the energy band derived
from $s$, $p_z$ and $\{p_x,p_y\}$ are symmetric in energy with
respect to their atomic energy band center, a feature usually referred to
as particle-hole symmetry, the interaction with the
higher lying Sb-$d$ bands here breaks this simplification. 

The important point is that the Fermi level
lies close to the intersection points of the $p_z$ derived  and \{$p_x,p_y\}$
derived Dirac cones, one of which lies a little above $E_F$ along $\Gamma-K$
and the other a little below $E_F$ along $\Gamma-M$. We can see that the bands
crossing at these points have different symmetry both in the flat and in the
slightly buckled case and are thus protected
by symmetry. Our symmetry analysis shows that it is a twofold rotation
along the $\Gamma-K$ direction which is
maintained both along $\Gamma-K$ and along $K-M$ even after buckling
and protects the existence of these band crossings.

The Dirac cones around $K$  are shown in Fig. \ref{fig:conesflat}(b-c). In the top row of the contour plots,
we present the DFT results and below it 
the corresponding results from an effective low energy Hamiltonian
derived based on symmetry and a nearest neighbor tight-binding model
described in full detail in Section IV of the SI. The conduction
band refers to the upward pointing $\{p_x,p_y\}$ derived cone in the center near $K$ while on the outside parts of the figure, beyond the cone intersection,
it refers to the downward pointing $p_z$-derived cone.
The opposite holds for the valence band.
We can see that the contours have a triangular shape
but are rotated by 30$^\circ$ from each other between the two cones.
This is also shown in a 3D view in part (c) and leads to a nodal line
with the Lissajous like shape. 
The triangular warping of the energy surfaces results from the
terms of order $q^2$ in an expansion around the point $K$ and can
be derived fully analytically from the tight-binding Hamiltonian
for the $p_z$ derived bands as is shown in the SI (Sec. IV). The linear terms
of the Dirac cones are isotropic. 
Because $p_z$ ($\{p_x,p_y\}$) orbitals are odd (even)
with respect to the horizontal mirror plane, they are derived
from a separate $2\times2$ and $4\times4$ Hamiltonian matrix. Both of these can
be further reduced to the eigenvalues of a $1\times1$ and $2\times2$
matrix because of the ``particle-hole'' symmetry within this model
and analytical expressions can be derived for them at $\Gamma$, $K$ and
near $K$. The threefold symmetry around $K$ is expected from the pointgroup
of $K$ which is $D_{3h}$. The Dirac cone states  can thus be written
\begin{eqnarray}
  E_z&=&\Delta_z\pm v_z q \pm \frac{q^2}{m_z}\cos{(3\phi)}, \nonumber \\
  E_\pi&=&\Delta_\pi\pm v_\pi q \pm \frac{q^2}{m_\pi}\cos{(3\phi)} \label{eqcone}
\end{eqnarray}
Here $\Delta_z$, $\Delta_\pi$ are the centers of the ($E^\dprime,E^\prime$)
Dirac cones at $K$, 
$v_z$, $v_\pi$ are the Dirac linear dispersion velocities a
and $m_z$, $m_\pi$ are effective mass parameters. The actual effective masses
depend on the direction of ${\bf q}$ represented by its
azimuthal angle $\phi$ from the $x$-axis ($\Gamma-K$ direction)
leading to a warping of the
constant energy lines with three-fold symmetry, while the velocity is
isotropic.  The opposite sign of the mass parameter for the
two cones leads to their relative rotation by 30$^\circ$ and is found
to be responsible for the interestingly shaped nodal line.  Within
the tight-binding model (SI-Sec.IV)
the sign of $m_\pi$ is found to be controlled
by the ratio of the $V_\sigma$ and $V_\pi$ interactions. 
The situation here is reminiscent of that in  AA bilayer graphene
but with the  difference that here the two Dirac cones have different
velocities and warping terms. As a result their intersection
is not a simple circular nodal line.

This unique shape of nodal ring gives rise to  an equally interestingly shaped
2D Fermi surface as shown in Fig. \ref{fig:conesflat}(d) (obtained from the effective mass Hamiltonian).
The Fermi surface can be seen to consist of electron and hole pockets
at 60$^\circ$ from each other and exhibits electron hole contact
points  (EHCP) where discontinuities occur in the band velocity.
This figure shows that the carrier type changes from electron to hole type
every 60$^\circ$ as we go around the Fermi surface,  an effect that has been named goniopolarity.\cite{Goldberger19} Furthermore, depending on the precise
location of the Fermi energy, which could in principle be varied by doping
or gating, the electron or hole transport could be larger  or smaller
as the direction is changed in-plane.
This situation is somewhat similar to the case of a  tilted nodal line
in a 3D {\bf k}-space, where interesting effects on the frequency dependent
conductivity
result from the cyclide geometry of the resulting Fermi surface.\cite{Ahn2017}
Here, we have calculated the static diffuse longitudinal
conductivity $\sigma(E_F,\phi)/\tau$ apart from the, at this point
unknown, relaxation time $\tau$ as function of azimuthal angle $\phi$
and the thermopower  (or Seebeck coefficient) which is proportional
to $d\ln{\sigma(E,\phi)}/dE|_{E=E_F}$ and whose sign reflects the charge
of the carriers.  These are shown in Fig. \ref{fig:conesflat}(f) and (g)
and show that the conductivity has modest anisotropy at the $\sim$14 \% level
but the thermopower changes discontinuously from positive to
negative at the EHCPs. Interestingly, it is positive for the directions
corresponding to electron transport because $d\sigma/dE|_{E_F}<0$.
This is because the conductivity varies
rapidly with energy near $E_F$ and the Fermi energy occurs just above
a peak in the $\sigma(E,\phi)$ related to a logarithmic singularity
in the density of states (DOS) resulting (see Fig.\ref{fig:conesflat}(e))
from the saddle-point  band structure at the point $B_{3g}$ at $M$. 
Inserting an order of magnitude estimate of $\tau=10^{-13}$s would
give a resistivity of order 0.2 $\mu\Omega$cm and a thermopower of
order 5 $\mu$V/K which is relatively high due to the Fermi level's
occurrence near a peak in the DOS.
The unique feature here however is the discontinuous angular
dependence of the thermopower. Numerous other opportunities
in the optical conductivity and magnetotransport 
related to this unique 2D nodal line remain to be explored. 

\begin{figure*}[!htb]
	\includegraphics[width=16cm]{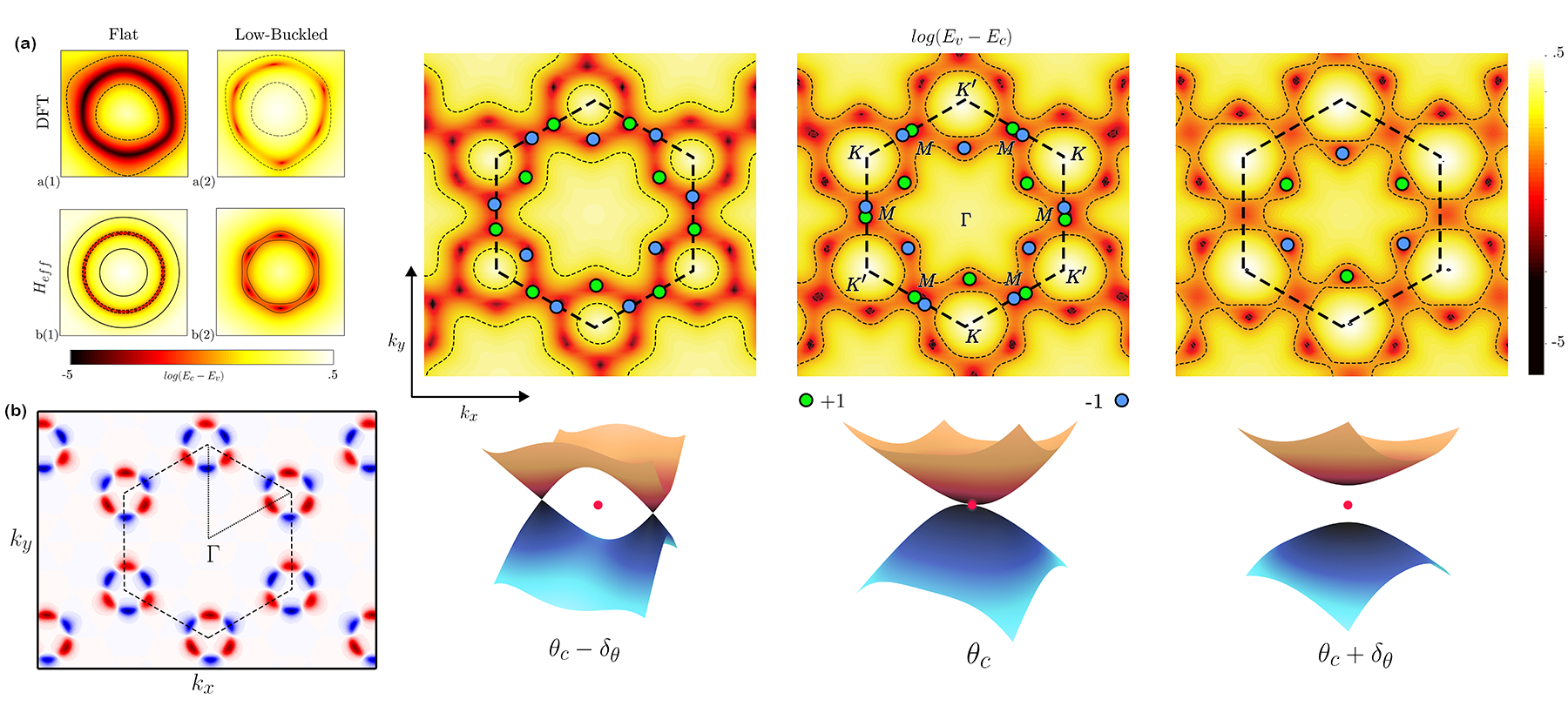}
	\caption{(a) Nodal line gap opening and formation of six Dirac points
          in slightly buckled honeycomb; (b) Berry flux around each of
          the Dirac cones surrounding each $K$-point; (c) movement and merging
          of Dirac cones as buckling angle increases; (d) corresponding
          behavior of constant energy surfaces around $M$ (red dot).\label{fig:flat-buckled}}
\end{figure*}

Next, we address the changes in band structure due to buckling.
The buckling leads to an interaction between the $p_z$ and
$\{p_x,p_y\}$ derived energy bands because the horizontal mirror plane
symmetry no longer applies. We assume here that the
bond-lengths between Sb atoms stays the same but the in-plane lattice
constant shrinks as the buckling is increased. This is qualitatively
consistent with the relaxation results\cite{Huang14} showing a decrease
in vertical distance between
the Sb atoms $d$ as function of in-plane lattice constant $a$.
Thus in our tight-binding
model the $V_\sigma$ and $V_\pi$ interactions stay the same but their
relative contribution to the hopping integrals changes. Within the
tight-binding model (SI-Sec. IV, Eqs.5-7),
the interaction terms between $p_z$ on the one hand
and $\{p_x,p_y\}$ on the other hand are proportional to $(V_\sigma-V_\pi)$
and to $\sin{(2\theta)}$ where $\theta$ is the buckling angle. For small
buckling the coupling is thus linear in $\theta$.  The symmetry labeled
band structure shows that in the six high symmetry directions around $K$
the crossing is protected by symmetry and thus the interaction
needs to go to zero every 60$^\circ$. Hence by symmetry, the low
energy Hamiltonian describing the behavior near these Dirac cones
can be written
\begin{equation}
  H^{buckled}_{{\bf K}+{\bf q}}=\begin{bmatrix}
\Delta_\pi +\frac{q^2\cos{(3\phi)}}{m_\pi} - q v_\pi& A\theta \sin{(3\phi)} \\ 
A \theta \sin{(3\phi)} &\Delta_z +\frac{q^2\cos{(3\phi)}}{m_z} + q v_z 
\end{bmatrix}
\label{eq:6}
\end{equation}
with $A$ some constant and a $\sin{(3\phi)}$ behavior of the off-diagonal
coupling.  \autoref{fig:flat-buckled} shows the effect of buckling on the nodal line around $K$ by plotting the difference between conduction and valence band in a log plot and verifies the existence of six Dirac points. 

Increasing the buckling either in the DFT or in the tight-binding model
we find that the Dirac touching points around $K$ move closer toward $M$
along the $K-M-K^\prime$ line and closer to $\Gamma$ along the $K-\Gamma$-line.
At some critical angle the two Dirac points along $K-K^\prime$ annihilate
each other when they reach $M$. This is shown in Fig.\ref{fig:flat-buckled}(b)
In the DFT results,  this occurs for
about $\theta_c\approx7^\circ$.

The reason why they can annihilate is
that they have opposite winding number $+1$ and $-1$.
The winding number was calculated
either from the effective low energy Hamiltonian or by calculating
the Berry curvature in the tight-binding model (SI, Sec.V).
The winding number, which can be thought of as a topological charge, is a conserved quantity \cite{Goerbig2017}. Thus the only possible way to remove Dirac cones is to merge and annihilate them. 
The reason for the merging of the Dirac cones is related to time reversal
symmetry, which 
guarantees that there is an equivalent and oppositely charged Dirac cone at $-{\bf k}$ for every Dirac cone at ${\bf k}$.
Thus by symmetry, these Dirac cones of opposite sign come together
and annihilate at the two types of time reversal invariant points
$\Gamma$ and $M$.

\citet{Montambaux2009} have analyzed the merging of 2D Dirac points
in terms of a universal Hamiltonian, which shows that near such a point
the bands correspond to a massive dispersion  in one of the in-plane
directions and
a massless one in the orthogonal in-plane direction, which
leads to an unusual $\sqrt{E}$ onset of the density of states.\cite{Hasegawa2006} which leads also to interesting changes in Landau levels\cite{Dietl2008}
The behavior of the energy surfaces near the merging is shown in Fig. \ref{fig:flat-buckled}(d).

After the merging of the Dirac points at $M$, upon further buckling, the
Dirac points along $K-\Gamma$ keep moving closer to $\Gamma$.  When 
they reach $\Gamma$ at a second critical angle of about 27$^\circ$, they
annihilate in pairs and the gap at $\Gamma$ beyond  this buckling becomes
non-trivial.  The evolution of the bands in the tight-binding model is
shown in more detail in SI. The $E^\prime$ and $E^\dprime$ points
splitting increases significantly with increasing buckling. With increasing
buckling the distinction between $p_z$ and $\{p_x,p_y\}$ becomes less
and less meaningful and a new type of hybridization between all three
bands forming a bonding set of bands and antibonding set of bands emerges
for the fully buckled ground state of the system. 
The opening of the gap
corresponds to the transition from a topologically non-trivial
to a trivial gap at $\Gamma$, which
was studied earlier in literature\cite{Huang14} starting from the equilibrium
large buckling by reducing the buckling under a tensile in-plane strain.

\begin{figure}[!htb]
	\includegraphics[width=8cm]{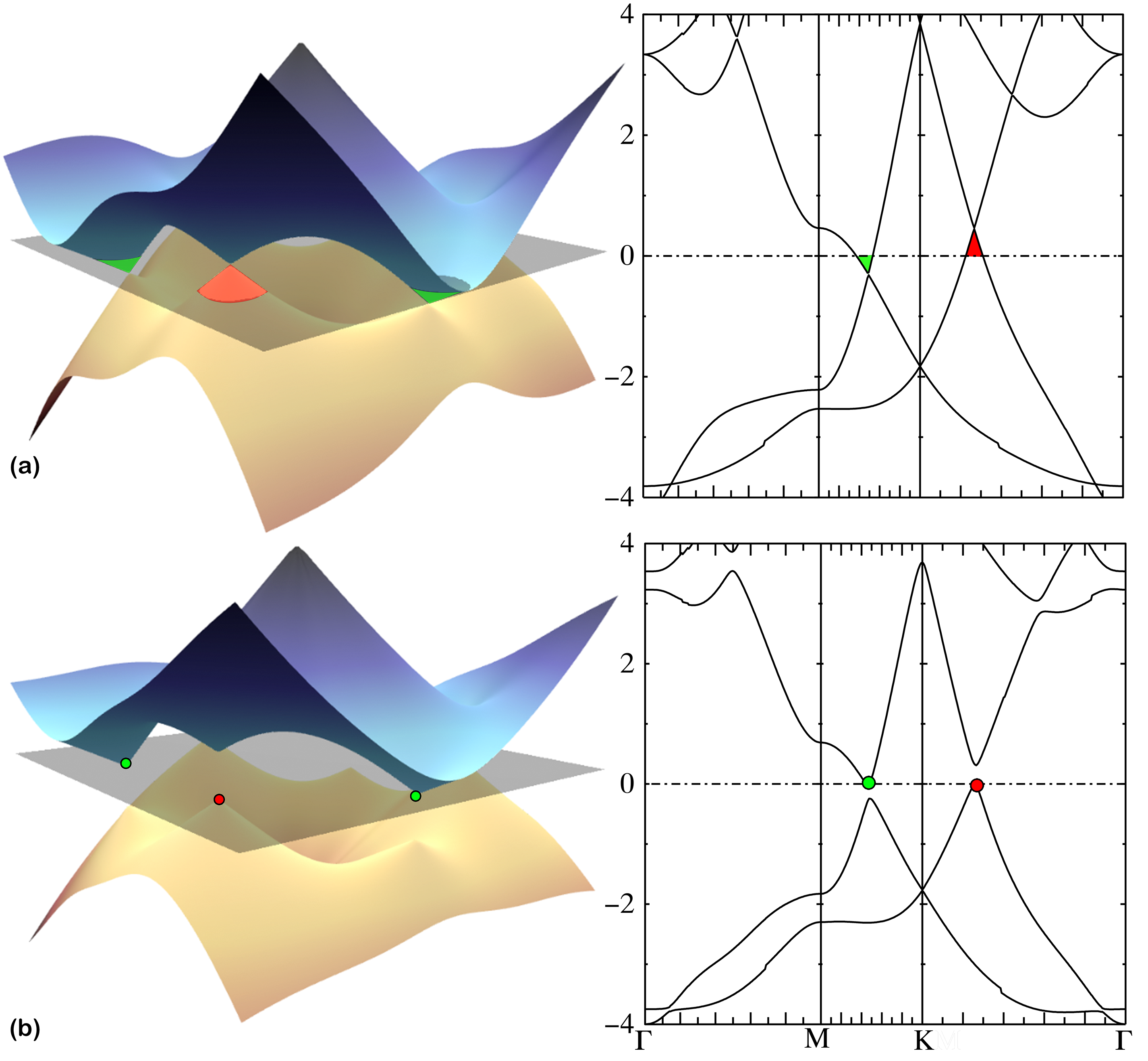}
	\caption{QS$GW$ band dispersion for the low-buckled system without
          (top) and with (bottom) SOC and the corresponding conical energy
        surfaces. Energies in eV. \label{fig:soc}}
\end{figure}

Turning on spin orbit coupling (SOC), a gap opens up at each of the 
Dirac points. We can see that it is larger for the upper $E^\prime$ point
(0.56 eV) than at the lower $E^\dprime$ point (0.17 eV) and intermediate
at the Dirac points near $E_F$.  The non-trivial nature of the
band crossings leads to topologically required edge states when the
gap is opened by SOC. For the nearly flat case, these were studied
in Ref. \onlinecite{Hsu2016}. 
However, we need to keep in mind how 
the gaps at the $K-M$ and $\Gamma-K$ Dirac points 
are placed energetically relative to each other.
While up to this point, we considered mostly generic properties which
are topologically invariant, we now need to worry about the accuracy
of the band structure and in particular the correct slope and placement
of the different Dirac points relative to each other.  To this
end we perform the band structures in the QS$GW$ approach which is known to give
much more accurate single particle excitations than DFT in a semilocal
approximation. 
We can see that this affects the band velocities of the Dirac cones
and the energy difference of the $E^\prime$, $E^\dprime$ Dirac points
at $K$. For the fully buckled equilibrium system QS$GW$ gives a gap of 2.9 eV,
significantly larger than the 1.3 eV obtained in GGA and somewhat larger
than the 2.28 eV from hybrid functional calculations.\cite{Zhang17}
In Fig. \ref{fig:soc} we can see that the highest
VBM at the Dirac point along $\Gamma-K$ lies at the same
energy as the lowest conduction band at the Dirac point along $K-M$.
So, the system is an indirect zero gap semiconductor. 
Because the SOC is weaker in arsenene, there is then a non-zero indirect overlap
between the occupied and empty bands at different Dirac points (see SI-Sec. VI).
The unique feature of
this band structure is that it should have  a topologically
spin polarized edge state associated with these
SOC induced spin-texture inverted gaps at specific ${\bf k}$-points,
even though the overall gap of the system is zero or slightly negative.
Such a situation has been labeled a gapless topological insulator (GTI).
It has been proposed to possibly occur due to electron-electron  interaction
effects\cite{Goncalves2019} but is here found even
in the independent particle approximation.

\begin{figure*}
  \includegraphics[width=18cm]{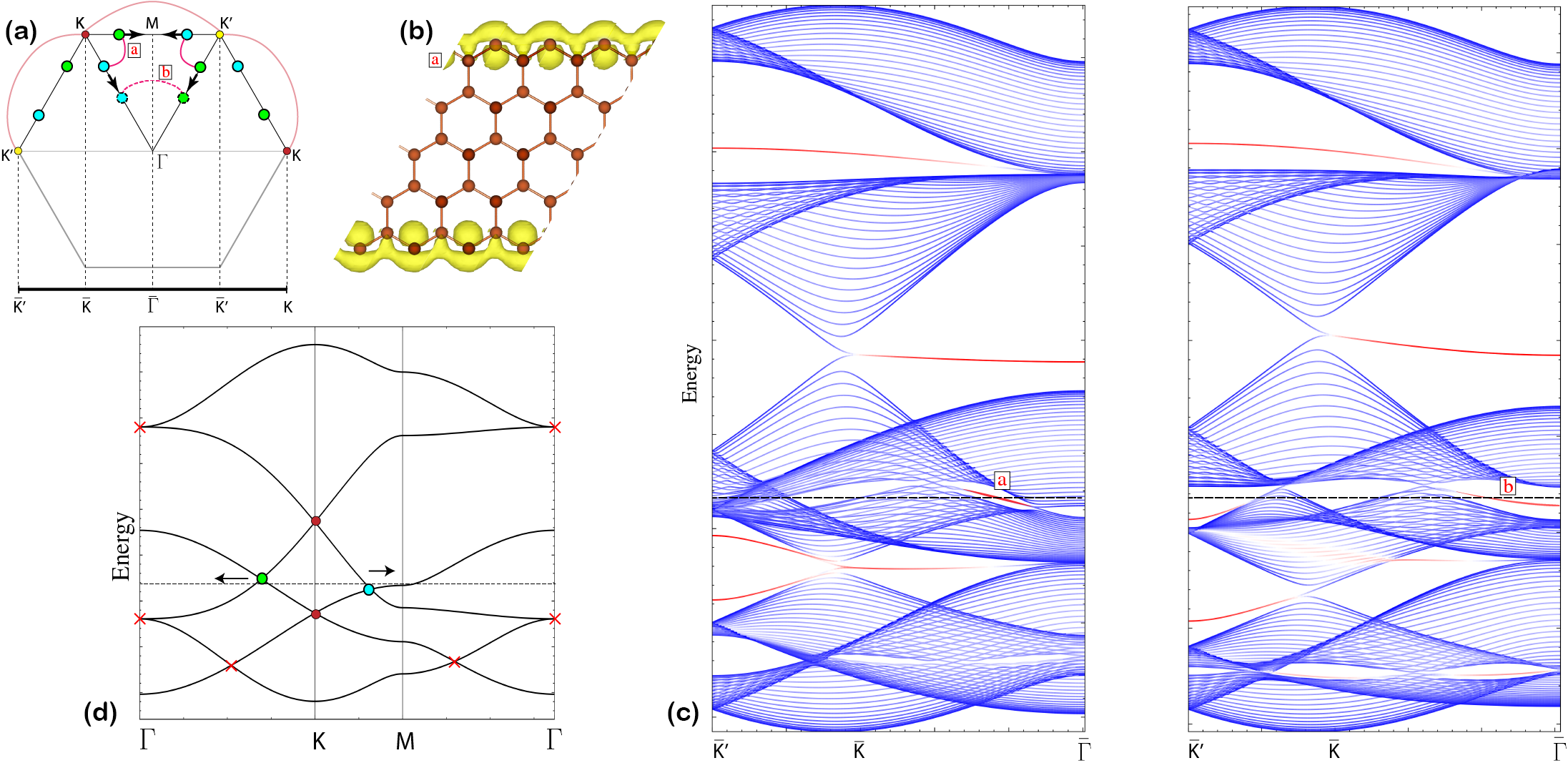}
  \caption{(a) folding of 2D Brillouin zone onto 1D Brillouin zone in nanoribbon
    indicating schematically the Dirac points linked by edge states
    and their motion under increased buckling; (b) 6 unit cell side nanoribbon with surface state from DFT; (c) band structure of
    nanoribbon in tight-binding model indicating the topologically
    protected edge states (in red) associated with Dirac point pairs labeled
    as in (a) for two buckling angles, one before and one after the merging of Dirac cones at $M$; (d) 2D tight-binding band structure for the low buckling case. \label{fig:edge}}
\end{figure*}

As is well known, topological features in the band structure are closely
related to protected edge states.  The tight-binding model
allows us to study the formation of these edge states explicitly in a finite
nanoribbon. We choose to cut the honeycomb lattice along the zigzag direction
and keep it periodic in the direction perpendicular to it but with a width
of 60 unit cells. As further confirmation,
Fig. \ref{fig:edge}(b) shows a smaller 6 unit cell nanoribbon,
calculated at the DFT level with the resulting surface state.
The Brillouin zone folding is shown
in Fig. \ref{fig:edge}(a). The Dirac points are color coded according to
their winding number and linked by lines for the pairs that will be connected
by corresponding edge states. These connections found here explicitly
from the numerical calculations
are consistent with the theoretical considerations
of Ryu and Hatsugai\cite{Ryu2002}.
The main part of the figure shows the 1D band structures
indicating the edge states in red labeled to identify them with particular
Dirac point pairings in the Brillouin zone figure following the same labeling. 
We here focus mainly
on the Dirac cones near the Fermi level but other ones can be seen
at energies farther away from the Fermi level.
These are related to other linear band crossings and the $\Gamma$-point
degenerate levels (marked by $\times$) which can be seen to occur in the tight-binding
band structure  for the 2D periodic band structure at energies farther removed
from the Fermi surface. In the low-buckling case,
from top to bottom, the edge states are related to
the $E_{1u}$ state at $\Gamma$, the $E^\prime$ state, the connection between
$\Gamma-K$ and $K-M$ Dirac points, the $E^\dprime$ Dirac point and the $E_{2g}$
state at $\Gamma$,  with the latter  two interacting along $\bar{K}-\bar{K^\prime}$. Finally, an edge state connected to the lowest energy Dirac crossings
along $\Gamma-K$ can be seen alon $\bar{K}-\bar{\Gamma}$.  The edge
states associated with the low energy crossing along $\Gamma-M$
cannot be seen in this nanoribbon because the $\Gamma-M$ direction is the one
along which we fold the bands. Because various of these edge states connect
Dirac points not along a high symmetry direction, one expects them
to be present for other cut-outs of the 2D lattice in arbitrary
directions.

In summary, in this paper we have shown that monolayer Sb and As
in the honeycomb structure exhibit  a rich
behavior in terms of topological features.
The system evolves from a unique type of nodal line in the flat case
to a series of six symmetry protected Dirac points surrounding each
$K$ point which move and annihilate first in pairs of opposite winding
number at a first critical buckling angle and subsequently
undergo a second topological transition when a trivial gap opens at $\Gamma$.
These result from the changing contribution of $\sigma$
and $\pi$ interactions between the orbitals as function of buckling.
The nodal line is here predicted to lead to a highly anisotropic in-plane
Seebeck coefficient reflecting goniopolarity.
A multitude of  edge states are predicted as well as a gapless topological
insulator behavior when spin-orbit coupling is included which
results from the energy straddling of the small gaps opening at
the symmetry protected Dirac states in the system. 

{\bf Acknowledgement}: This work was supported by the U.S. Department of
Energy-Basic Energy Sciences under grant No. DE-SC0008933.  The calculations made use of the High Performance Computing Resource in the Core Facility for Advanced Research Computing at Case Western Reserve University.

{\bf Methods}:
    The calculations of the structural  stability and band structure
    were performed using density functional theory in the Perdew-Becke-Ernzerhof (PBE)\cite{PBE} generalized    gradient approximation (GGA).
    Band structures were also calculated using the quasiparticle self-consistent
    QS$GW$  many-body perturbation theory method.\cite{MvSQSGWprl,kotani:QSGW}
    Here $GW$ stands
    for the one-electron Green's function and $W$ for the screened
    Coulomb interaction.\cite{Hedin65,Hedin69}
    All calculations were performed
    using the full-potential
    linearized muffin-tin orbital (FP-LMTO) method\cite{Methfessel,Kotani10}
    using the questaal package, which is fully described in Ref.\cite{questaalpaper} and available at \cite{questaal}. Convergence parameters were chosen
    as follows: basis set $spdf-spd$ spherical wave envelope functions  plus augmented plane waves with a cut-off of 3 Ry, augmentation cutoff $l_{max}=4$,
    {\bf k}-point mesh, $12\times12\times2$. The monolayer slabs were
    separated by a vacuum region of 3 nm.  In the $GW$ calculations  the self
    energy $\Sigma$ is calculated  on a {\bf k}-mesh of $5\times5\times2$
    points and interpolated to the above finer mesh and the bands along
    symmetry lines using the real space representation of the LMTO basis set. 
    The diffusive conductivity tensor $\sigma/\tau$ (apart from the unknown
    relaxation time) 
    was calculated using the equation
    \begin{equation}
      \sigma_{\alpha\beta}(E)/\tau=e^2\sum_n\int \frac{d^3k}{4\pi^3}
      \delta(E-E_n({\bf k})) v_\alpha v_\beta
    \end{equation}
    with $v_\alpha=\partial E_n({\bf k})/\partial k_\alpha$ The conductivity and
    Seebeck coefficients are then obtained from
    \begin{eqnarray}
      \sigma_{\alpha\beta}/\tau&=&e^2\int dE \sigma_{\alpha\beta}\left(-\frac{\partial f}{\partial\mu}\right)\approx \sigma_{\alpha\beta}(\mu) \nonumber \\
      S_{\alpha\beta}&=&-\left.\frac{\pi^2 k_B^2T}{e}\frac{d\ln{\sigma_{\alpha\beta}(E)}}{dE}\right|_{E=\mu}
    \end{eqnarray}
    with $\mu=E_F$ the Fermi energy. 
    The nearest neighbor tight-binding model used to analyze the results
    and to obtain the nanoribbon surface states is described in SI.

\bibliography{sb.bib,dft.bib,gw.bib,lmto.bib}
   \section{Supplemental Information}
    \subsection{Introduction}
    In this Supplemental Information we provide various details and further
    information supporting our main paper. They are presented in the order
    they appear in the main paper. First, a discussion of the metastability
    of the flat structure compared to the equilibrium buckled structure
    is presented in Sec. \ref{sec:meta}. Then the symmetry labeling
    analysis is presented in Sec. \ref{sec:sym}. Next, the tight-binding model
    and its analysis are presented in Sec.\ref{sec:tb}. The calculation
    of the winding number is explained in Sec. \ref{sec:winding}. 
    Finally, in Sec. \ref{sec:As}
    we give a few corresponding results for As in comparison with the main
    results on Sb in the main paper.

    \subsection{Stability and Metastability of flat and buckled 2D monolayers}
    \label{sec:meta}
\begin{figure}[!htb]
    	\includegraphics[width=9cm]{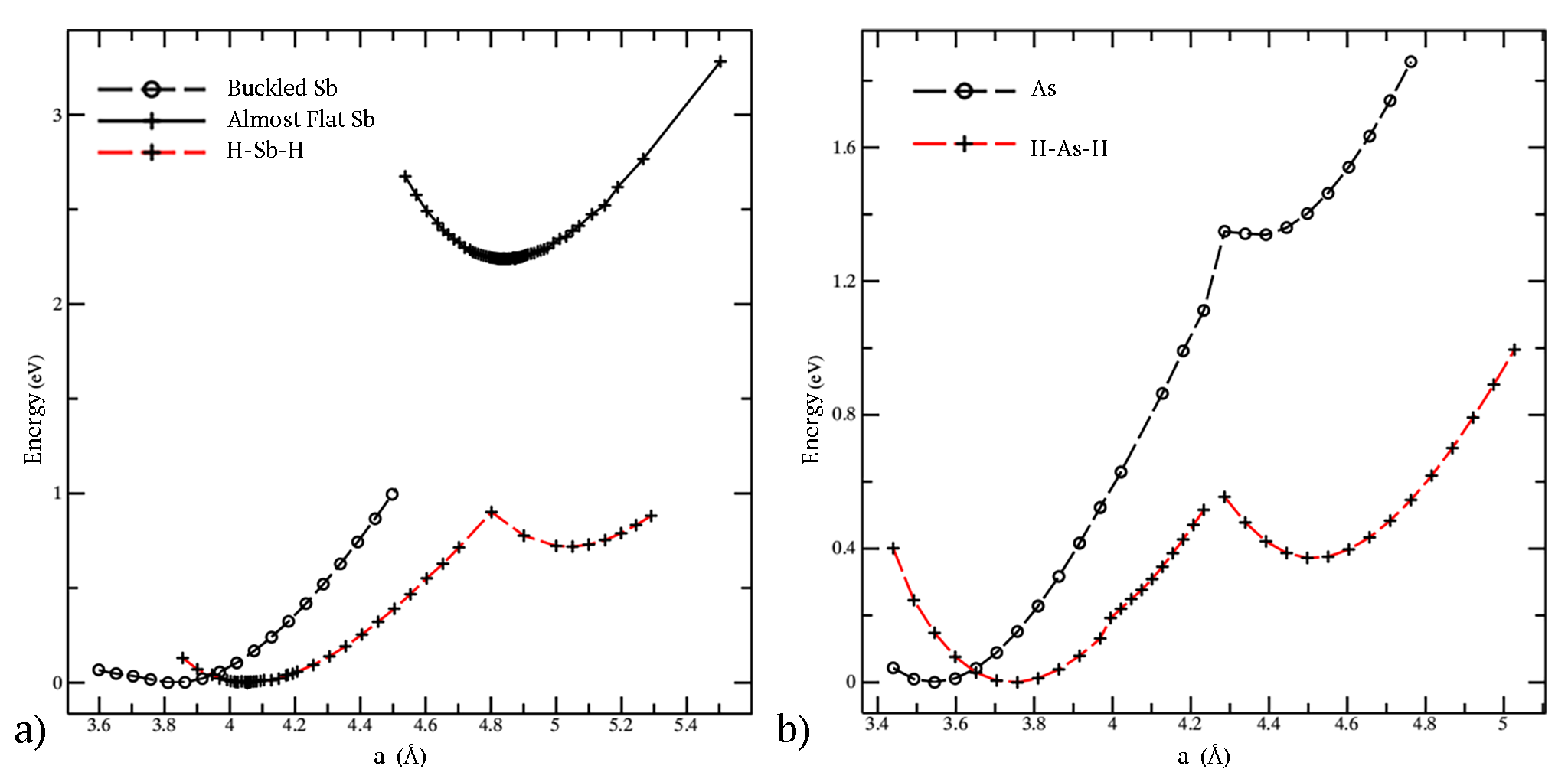}
    	\caption{Fully relaxed total energy
    		per 2-atom unit cell as function of in-plane
    		lattice constants for Sb (left) and As (right).
    		\label{fig:tote}}
    \end{figure}

    Here we discuss the stability of flat {\sl vs.} buckled
    forms of 2D monolayer Sb and As in the $\beta$-structure.
    In Fig. \ref{fig:tote} we show the results of first-principles density functional calculations
    using the PBE-GGA functional for the total energy as function of in-plane
    lattice constant of the honeycomb lattice.
    In the buckled case, the
    structure for a given $a$ is allowed to fully relax, leading to a
    relative high buckling angle of  $\sim33^\circ$
    defined by $\tan{\theta}=\sqrt{3}d/a$
    with $d$ the vertical distance between  Sb atoms along the $z$-axis
    and $a$ the in plane lattice constant. 
    We can see that a nearly flat structure, with higher in-plane
    lattice constant, exists as  a second metastable minimum. 
    
    Finally hydrogenating the structure with H on top or below the Sb on
    alternating Sb, is seen to lower the energy of the flat
    Sb structure significantly although it still has higher energy than
    the equilibrium structure.   Similar results are found for As.
    In the main paper we have not studied the hydrogenated form.
    In that case the Fermi level lies at the $\{p_x,p_y\}$-derived
    $E^\prime$ Dirac point at $K$. This case was studied in
    Refs. \cite{Gong18,WangYaping2016}
    
\begin{figure}
    	\includegraphics[width=9 cm]{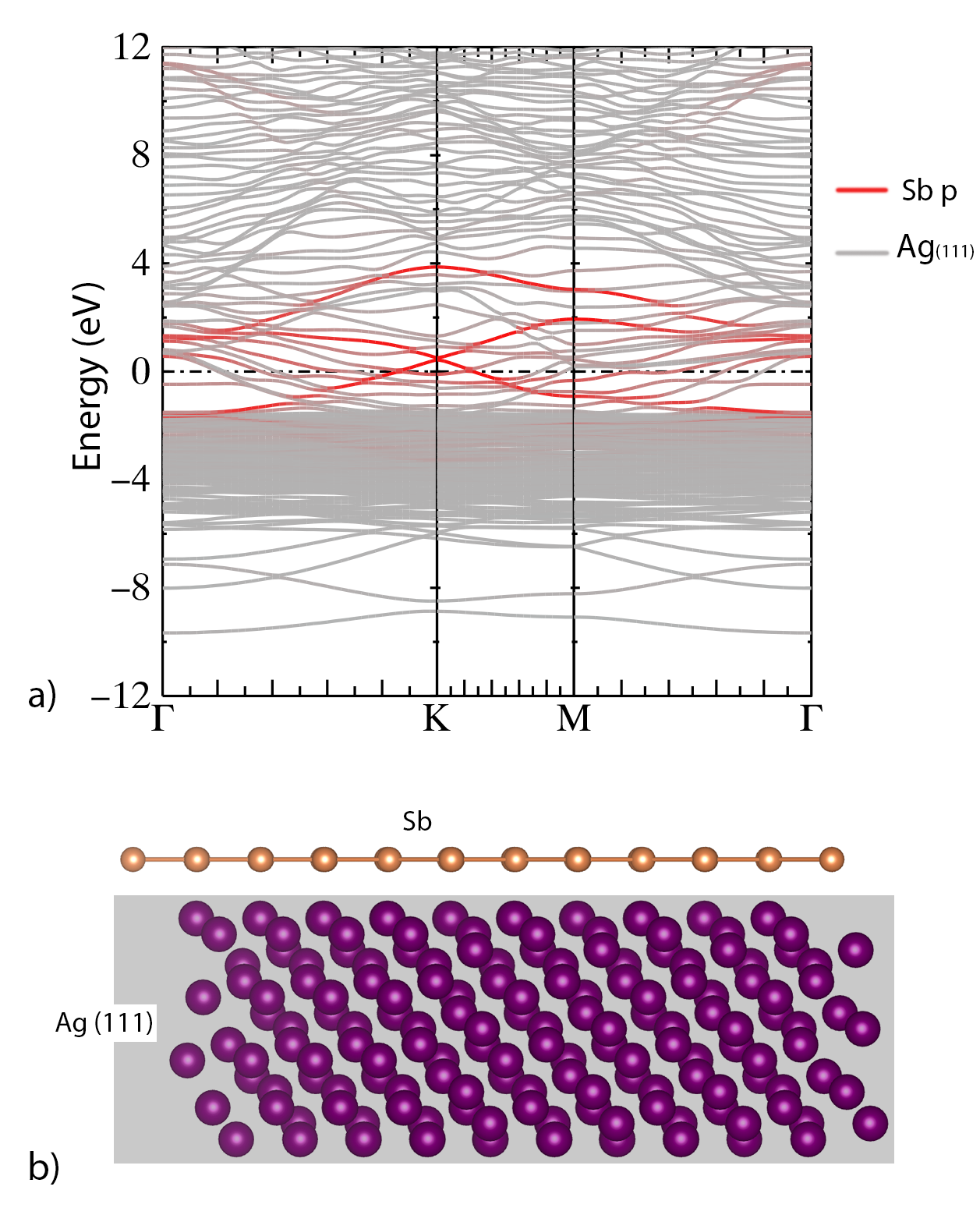}
    	\caption{Band structure of a Ag (111)  slab with Sb monolayer
    		adsorbed.\label{fig:bndsagsb}}
    	
    \end{figure}

    Although free standing monolayer Sb in the $\beta$-structure clearly
    has a high buckling angle, the flat or nearly flat structure can be
    stabilized by epitaxial in-plane tensile strain by putting the Sb on
    a Ag(111) structure.\cite{Shao18} The band structure of free standing flat monolayer Sb  is discussed in the main paper.
    For comparison, we here show the band structure
    of a 10 layers thick
    Ag layer with a monolayer of Sb on top in Fig. \ref{fig:bndsagsb}.
    The structure was relaxed with DFT before calculating the band structure.  
    The bands weighted by their
    Sb contribution are shown in red. This shows that the features of
    monolayer Sb can still be recognized clearly on top of the Ag background,
    especially in the important region near the Dirac crossings and Fermi energy
    where the Ag density of states is low. 
    This shows that there is a route forward to experimentally investigate the
    band structure aspects studied in the main paper
    by means of epitaxial stabilization
    and the investigation of monolayer Sb under high tensile strain
    in a flat or nearly flat form is not just a theoretical exercise
    but could potentially be realized experimentally by adjusting
    the coupling to the underlying substrate or varying the lattice constant
    of the substrate. Of course, for  topologically induced effects on the
    transport, one would then also have to consider scattering to and from the
    underlying Ag band structure. 
    
    \subsection{Symmetry Analysis} \label{sec:sym}
    Here we present the details of the symmetry labeling of the
    band structure. 
    Flat monolayer Sb in the honeycomb structure has the same spacegroup and pointgroup as graphene. Several prior papers have addressed the symmetry labeling of
    the bands in graphene\cite{Kogan12,Kogan14,Minami}
    but still some confusion exists because of the
    non-uniqueness in specifying the symmetry operations  and irreducible representations.
    The point group of the crystal is $D_{6h}$ and this is also the point group at $\Gamma$.
    The character table is given in Table \ref{tabD6h}. Note that in
    $D_{6h}$ there are two sets of 2-fold rotations lying in the plane.
    We choose $U_2$ to pass through the atoms, while the $U_2^\prime$ pass
    through the bond centers.  In reciprocal space this implies that $U_2$ lies along the $\Gamma-M$ lines and $U_2^\prime$ lies along the $\Gamma-K$ lines.
    The corresponding mirror planes $i*U_2=\sigma_v$ and $i*U_2^\prime=\sigma_v^\prime$ are perpendicular to these axes, so $\sigma_v$ goes through the bond centers
    and $\sigma_v^\prime$ goes through the atoms. We choose the lattice vectors as ${\bf a}_1=a\hat{\bf x}$,
    ${\bf a}_2=-\frac{1}{2}a\hat{\bf x}+\frac{\sqrt{3}}{2}a\hat{\bf y}$ and one of the points $K$ in the Brillouin zone
    thus lies along $x$ and one of the points $M$ lies along $y$. 
    The $K$-points rotated by 60$^\circ$ we label $K^\prime$. We note points $K$  rotated by 120$^\circ$ are equivalent
    in that they are related by a reciprocal lattice vector whereas $K$ and $K^\prime$ are not.
    Likewise we denote the $M$-points rotated by 60$^\circ$, $M^\prime$ and rotated by 120$^\circ$  as $M^\dprime$.
    These are nonequivalent but rotating them by 180$^\circ$ gives equivalent $M$-points. 
    
    The group $D_{6h}$ can be viewed as the direct product $D_6\otimes C_i$ where $C_i$ is the group consisting
    of the identity and the inversion. (It could also be viewed as $C_{6v}\otimes C_i$ or $C_{6v}\otimes C_s$
    or $D_6\otimes C_s$ with $C_s$ the group formed by $\{E,\sigma_h\}$ and this is the reason behind
    some of the discrepancies between previous symmetry labelings.)
    The irreducible representations follow the usual notation in which subscript $g$ means even with
    respect to inversion and $u$ means odd with respect to inversion. The corresponding 
    labels of the Koster notation\cite{Koster} are also included. 
    \begin{table}
    	\caption{Character table of $D_{6h}$\label{tabD6h}}
    	\begin{ruledtabular}
    		\begin{tabular}{ll|rrrrrr|rrrrrr}
    			$D_{6h}$ & & $E$ & $C_2$ & $2C_3$ & $2C_6$ & $3U_2$& $3U_2^\prime$ & $i$ &
    			$\sigma_h$ & $2S_6$ & $2S_3$ & $3\sigma_v$ & $3\sigma_v^\prime$ \\ \hline
    			$\Gamma_1^+$ & $A_{1g}$ & 1 & 1 & 1 & 1 & 1 & 1 &  1 & 1 & 1 & 1 & 1 & 1 \\
    			$\Gamma_2^+$  & $A_{2g}$ & 1 & 1 & 1 & 1 & $-1$ & $-1$ &  1 & 1 & 1 & 1 & $-1$ & $-1$ \\
    			$\Gamma_4^+$ & $B_{1g}$  & 1 & $-1$ & 1 & $-1$ & 1& $-1$ &  1 & $-1$ & 1 & $-1$ & 1& $-1$ \\
    			$\Gamma_3^+$ & $B_{2g}$  & 1 & $-1$ & 1 & $-1$ & $-1$ & 1 &   1 & $-1$ & 1 & $-1$ & $-1$ & 1 \\
    			$\Gamma_6^+$ & $E_{2g}$  & 2 & 2 & $-1$ & 1 & 0 & 0 &  2 & 2 & $-1$ & 1 & 0 & 0 \\
    			$\Gamma_5^+$ & $E_{1g}$  & 2 & $-2$ & $-1$ & 1& 0 & 0 &    2 & $-2$ & $-1$ & 1& 0 & 0  \\ \hline
    			$\Gamma_1^-$ & $A_{1u}$  & 1 & 1 & 1 & 1 & 1 & 1 & $-1$ & $-1$ & $-1$ & $-1$ & $-1$ & $-1$ \\
    			$\Gamma_2^-$ & $A_{2u}$  & 1 & 1 & 1 & 1 & $-1$ & $-1$ & $-1$ & $-1$ & $-1$ & $-1$ & 1 & 1 \\
    			$\Gamma_4^-$ & $B_{1u}$  & 1 & $-1$ & 1 & $-1$ & 1& $-1$ &  $-1$ & 1 & $-1$ & 1& $-1$ & 1 \\
    			$\Gamma_3^-$ & $B_{2u}$  & 1 & $-1$ & 1 & $-1$ & $-1$ & 1 & $-1$ & 1 & $-1$ & 1 & 1 & $-1$ \\
    			$\Gamma_6^-$ & $E_{2u}$  & 2 & 2 & $-1$ & 1 & 0 & 0 & $-2$ & $-2$ & 1 & $-1$ & 0 & 0 \\
    			$\Gamma_5^-$ & $E_{1u}$  & 2 & $-2$ & $-1$ & 1 & 0 & 0 & $-2$ & 2 & 1 & $-1$ & 0 & 0 \\ 
    		\end{tabular}
    	\end{ruledtabular}
    \end{table}
    Although the symmetry aspects for graphene and Sb are the same, a difference is that the $s$-$p$ splitting
    is larger in Sb relative to the hopping interactions and hence the $s$ states form separate bands
    from the atomic $p$-state derived bands. 
    The $s$-states are even in the horizontal mirror plane and form bonding an antibonding combinations
    on atoms $A$ and $B$ in the unit cell: $(s_A+s_B)/\sqrt{2}$ and $(s_A-s_B)/\sqrt{2}$. The operations
    $C_2$, $C_6$, $U_2^\prime$, $i$, $S_6$, $\sigma_v$ change sublattices $A$ to $B$ and vice versa.  
    Thus it is clear that the bonding state belongs to $A_{1g}$ and the antibonding state to $B_{1u}$.
    The $p_z$ states are odd {\sl vs.} the $\sigma_h$ and are thus decoupled by symmetry from the $s$ and $p_x,p_y$.
    Again they form bonding and antibonding states $(p_{zA}+p_{zB})/\sqrt{2}$ and $(p_{zA}-p_{zB})/\sqrt{2}$
    which are now respectively of symmetry $A_{2u}$ and $B_{2g}$.  The $p_x$ and $p_y$ build the representations
    $E_{2g}$ and $E_{1u}$.
    
    Now, the group of {\bf k}, ${\cal G}_{\bf k}$, consists of those operations that turn {\bf k} into
    itself up to a reciprocal lattice vector. For point $K$, which we choose along the $x$-axis,
    these are $\{E,2C_3,3U_2^\prime,\sigma_h, 2S_3,3\sigma_v$\}, which build the group $D_{3h}$. At $M$
    the group ${\cal G}_{{\bf k}_M}$ consists of $\{E,C_2, U_2,U_2^\prime,i,\sigma_h,\sigma_v,\sigma_v^\prime\}$ building the group $D_{2h}$.  The character tables of these groups are given in Tables \ref{tabD3h}, \ref{tabD2h}.
    \begin{table}
    	\caption{Character table of $D_{3h}$.\label{tabD3h}}
    	\begin{ruledtabular}
    		\begin{tabular}{ll|rrr|rrr}
    			$D_{3h}$ & & $E$ & $2C_3$ & $3U_2^\prime$ & $\sigma_h$ & $2S_3$ & $3\sigma_v$ \\ \hline
    			$K_1$ & $A_1^\prime$ & 1 & 1 & 1 & 1 & 1 & 1 \\
    			$K_2$ & $A_2^\prime$ & 1 & 1 & $-1$ &  1 & 1 & $-1$ \\
    			$K_6$ & $E^\prime$   & 2 & $-1$ & 0& $2$ & $-1$ & 0 \\ \hline 
    			$K_3$ & $A_1^\dprime$ & 1 & 1 & 1 & $-1$ & $-1$ & $-1$ \\
    			$K_4$ & $A_2^\dprime$ & 1 & 1 & $-1$ & $-1$ & $-1$ & 1 \\
    			$K_5$ & $E^\dprime$   & 2 & $-1$ & 0 & $-2$ & 1 & 0 \\
    		\end{tabular}
    	\end{ruledtabular}
    \end{table}
    Note that in the group $D_{3h}$ representations labeled by superscript $^\prime$ are even with respect to the
    $\sigma_h$ and $^\dprime$ are odd. In $D_{2h}$ note that which irrep is called $B_{2g}$ or $B_{3g}$ depends
    on the choice of $U_2$ or $U_2^\prime$ being chosen as first or second set of 2-fold axes in the $xy$-plane.
    We added another label to indicate specifically which axes are chosen for the $M$ lying along $y$.
    
    \begin{table}
    	\caption{Character table of $D_{2h}$.\label{tabD2h}}
    	\begin{ruledtabular}
    		\begin{tabular}{ll|rrrr|rrrr}
    			$D_{2h}$ && $E$ & $C2$ & $U_{2(y)}$ & $U^\prime_{2(x)}$ & $i$ & $\sigma_h$ & $\sigma_{v(xz)}$ & $\sigma_{v(yz)}^\prime$ \\ \hline
    			$M_1^+$ & $A_g$  & 1 & 1 & 1 & 1 & 1 &  1 & 1 & 1  \\
    			$M_3^+$ & $B_{1g}$ & 1 & 1 & $-1$ & $-1$ &  1 & 1 & $-1$ & $-1$  \\
    			$M_2^+$ & $B_{2g}$ & 1 & $-1$ & 1 & $-1$ &  1 & $-1$ & 1 & $-1$ \\
    			$M_4^+$ & $B_{3g}$ & 1 & $-1$ & $-1$ & 1 &  1 & $-1$ & $-1$ & 1 \\ \hline
    			$M_1^-$ & $A_u$    & 1 & 1 & 1 & 1  & $-1$ & $-1$ & $-1$ & $-1$ \\
    			$M_3^-$ & $B_{1u}$ &  1 & 1 & $-1$ & $-1$ & $-1$ & $-1$ & 1 & 1 \\
    			$M_2^-$ & $B_{2u}$ &  1 & $-1$ & 1 & $-1$ & $-1$ & 1 & $-1$ & 1 \\
    			$M_4^-$ & $B_{3u}$ &   1 & $-1$ & $-1$ & 1 & $-1$ & 1 & 1 & $-1$ \\
    		\end{tabular}
    	\end{ruledtabular}
    \end{table}
    Along the $\Gamma-K=T$ axis the symmetry operations remaining are $\{E,U^\prime_{2(x)}, \sigma_h,\sigma_{v(xz)}\}$
    building the group $C_{2v}$.  The irreps of this group depend on which of the mirror planes one chooses
    as first. We here choose $\sigma_h$ as first mirror plane. Along $\Gamma-M=\Sigma$ the group is also
    $C_{2v}$ but now the symmetry elements remaining are $\{E,U_{2(y)},\sigma_h,\sigma_{v(yz)}^\prime\}$.
    Finally along the line $M-K$ the group is the same as along $\Gamma-K$.
    The character tables used are given in Table \ref{tabC2v}.
    \begin{table}
    	\caption{Character table of $C_{2v}$.\label{tabC2v}}
    	\begin{ruledtabular}
    		\begin{tabular}{ll|rrrr}
    			$C_{2v}$ & $\Gamma-K$  & $E$ & $U^\prime_{2(x)}$ & $\sigma_h$ & $\sigma_{v(xz)}$ \\
    			& $\Gamma-M$  & $E$ & $U_{2(y)}$       & $\sigma_h$ & $\sigma^\prime_{v(yz)}$ \\ \hline
    			$(T,\Sigma)_1$   & $A_1$ & 1 & 1 & 1 & 1 \\
    			$(T,\Sigma)_3$   & $A_2$ & 1 & 1 & $-1$ & $-1$ \\
    			$(T,\Sigma)_2$   & $B_1$ & 1 & $-1$ & 1 & $-1$ \\
    			$(T,\Sigma)_4$   & $B_2$ & 1 & $-1$ & $-1$ & 1 \\
    		\end{tabular}
    	\end{ruledtabular}
    \end{table}
    
    \begin{table*}[]
    	\centering
    	\caption{Groups ${\cal G}_{\bf k}$ for flat and buckled cases}
    	\begin{ruledtabular}
    		\begin{tabular}{@{}lllllll@{}}
    			\toprule
    			& $\Gamma$                      & $\Gamma$-M                                                           & M                             & M-K                                                                  & K                             & K-$\Gamma$                                      \\ \midrule\hline
    			\multicolumn{1}{l|}{\multirow{2}{*}{Flat}}    & \multicolumn{1}{l|}{$D_{6h}$} & \multicolumn{1}{l|}{$C_{2v}$}                                        & \multicolumn{1}{l|}{$D_{2h}$} & \multicolumn{1}{l|}{$C_{2v}$}                                        & \multicolumn{1}{l|}{$D_{3h}$} & $C_{2v}$                                        \\
    			\multicolumn{1}{l|}{}                         & \multicolumn{1}{l|}{}         & \multicolumn{1}{l|}{$\{E,U_{2(y)},\sigma_h,\sigma^\prime_{v(yz)}\}$} & \multicolumn{1}{l|}{}         & \multicolumn{1}{l|}{$\{E,U^\prime_{2(x)},\sigma_h,\sigma_{v(xz)}\}$} & \multicolumn{1}{l|}{}         & $\{E,U^\prime_{2(x)},\sigma_h,\sigma_{v(xz)}\}$ \\ \midrule\hline
    			\multicolumn{1}{l|}{\multirow{2}{*}{Buckled}} & \multicolumn{1}{l|}{$D_{3d}$} & \multicolumn{1}{l|}{$C_s$}                                           & \multicolumn{1}{l|}{$C_{2h}$} & \multicolumn{1}{l|}{$C_2$}                                           & \multicolumn{1}{l|}{$D_3$}    & $C_2$                                           \\
    			\multicolumn{1}{l|}{}                         & \multicolumn{1}{l|}{}         & \multicolumn{1}{l|}{$\{E,\sigma^\prime_{v(yz)}\}$}                   & \multicolumn{1}{l|}{}         & \multicolumn{1}{l|}{$\{E,U^\prime_{2(x)}\}$}                         & \multicolumn{1}{l|}{}         & $\{E,U^\prime_{2(x)}\}$                         \\ \bottomrule
    		\end{tabular}
    		
    		\label{tab:sym-label}
    	\end{ruledtabular}
    \end{table*}
    
    Next, we consider the modifications that occur upon buckling.  In this case, the mirror planes through  the atoms
    remain but the 2-fold rotation through the atoms is no longer valid.  Similarly, the rotation axis through the bond
    center remains but the mirror plane through the bond axis is no longer valid. The six-fold rotations also are no longer valid
    but the inversion remains.  The corresponding changes from flat
    to buckled in the groups ${\cal G}_{\bf k}$ are given in
    Table \ref{tab:sym-label}. 
    The resulting group is $D_{3d}$ and its character table is given in Table \ref{tabD3d}.
    \begin{table}
    	\caption{Character table for $D_{3d}$.\label{tabD3d}}
    	\begin{ruledtabular}
    		\begin{tabular}{ll|rrr|rrr}
    			$D_{3d}$ & & $E$ & $2C_3$ & $3U_2^\prime$ & $i$ & $2S_6$ & $3\sigma_v^\prime$ \\ \hline
    			$\Gamma_1^+$ & $A_{1g}$ & 1 & 1 & 1 & 1 & 1 & 1 \\
    			$\Gamma_2^+$ & $A_{2g}$ & 1 & 1 & $-1$ &  1 & 1 & $-1$ \\
    			$\Gamma_3^+$ & $E_g$    & 2 & $-1$ & 0 &  2 & $-1$ & 0 \\ \hline
    			$\Gamma_1^-$ & $A_{1u}$ & 1 & 1 & 1 & $-1$ & $-1$ $-1$ \\
    			$\Gamma_2^-$ & $A_{2u}$ & 1 & 1 & $-1$ & $-1$ & $-1$ & 1 \\
    			$\Gamma_3^-$ & $E_u$   & 2 & $-1$ & 0  & $-2$ & 1 & 0 \\
    		\end{tabular}
    	\end{ruledtabular}
    \end{table}
    
    We can thus easily convert the labels from the $D_{6h}$ case to the $D_{3d}$ case according to
    Table \ref{compat}. The same holds for the odd {\sl vs.} inversion irreps labeled by the $u$ subscript. 
    \begin{table}
    	\caption{Compatibility between $D_{6h}$ and $D_{3d}$ group irreps.\label{compat}}
    	\begin{tabular}{l|cccccc} \hline
    		$D_{6h}$ & $A_{1g}$ & $A_{2g}$ & $B_{1g}$ & $B_{2g}$ & $E_{2g}$ & $E_{1g}$ \\
    		$D_{3d}$ & $A_{1g}$ & $A_{2g}$ & $A_{2g}$ & $A_{1g}$ & $E_g$   & $E_g$ \\ \hline
    	\end{tabular}
    \end{table}
    
    Now at $K$, the group becomes $D_3$  because we loose the inversion. The character table
    thus consist just of the upper left block in Table \ref{tabD3d}. The irreps stay
    the same as at $\Gamma$ but without the $g,u$ subscripts. At $M$ we now have the group $C_{2h}$
    consisting of $\{E,U^\prime_{2(x)},i,\sigma^\prime_{v(yz)}\}$.
    Its character table is given in Table \ref{tabC2h}.
    
    \begin{table}
    	\caption{Character table of $C_{2h}$.\label{tabC2h}}
    	\begin{ruledtabular}
    		\begin{tabular}{ll|rr|rr}
    			$C_{2h}$ & & $E$ & $U^\prime_{2(x)}$ & $i$ & $\sigma^\prime_{v(yz)}$ \\ \hline
    			$M_1^+$  & $A_g$ & 1 & 1 & 1 & 1 \\
    			$M_2^+$  & $B_g$ & 1 & $-1$ & 1 & $-1$ \\ \hline
    			$M_1^-$  & $A_u$ & 1 & 1 & $-1$ & $-1$ \\
    			$M_2^-$  & $B_u$ & 1 & $-1$ & $-1$ & 1 \\
    		\end{tabular}
    	\end{ruledtabular}
    \end{table}
    
    Along the line $\Gamma-K$ the group is $C_2$ consisting of $\{E, U^\prime_{2(x)}\}$.
    The same applies to the line $M-K$.  The character table is given in Table \ref{tabC2}.
    Along $\Gamma-M$, the group is $C_s$ consisting of $\{E,\sigma^\prime_{v(yz)}\}$
    with characters given in Table \ref{tabCs}.
    
    \begin{table}
    	\caption{Character table of $C_2$.\label{tabC2}}
    	\begin{ruledtabular}
    		\begin{tabular}{ll|rr}
    			$C_2$ & & $E$ & $U^\prime_{2(x)}$ \\ \hline
    			$T_1$ & $A$ & 1 & 1 \\
    			$T_2$ & $B$ & 1 & $-1$
    		\end{tabular}
    	\end{ruledtabular}
    \end{table}
    
    \begin{table}
    	\caption{Character table of $C_s$.\label{tabCs}}
    	\begin{ruledtabular}
    		\begin{tabular}{ll|rr}
    			$C_s$ & & $E$ & $\sigma^\prime_{v(yz)}$ \\ \hline
    			$\Sigma_1$ & $A^\prime$ & 1 & 1 \\
    			$\Sigma_2$ & $A^\dprime$ & 1 & $-1$ \\
    		\end{tabular}
    	\end{ruledtabular}
    \end{table}
    
    We may also consider a flat structure but making the $A$ and $B$ atoms different. This would apply to the case of monolayer h-BN. 
    Then starting from $D_{6h}$ we loose the inversion but we keep the horizontal mirror plane.
    The group at $\Gamma$ in that case is $D_{3h}$ consisting of $\{E,2C_3,3U_2,\sigma_h,2S_3, 3\sigma_v^\prime\}$.
    The group at $K$ in that case is the group $C_{3h}$ consisting of $\{ E, 2C_3, \sigma_h, 2S_3\}$.
    At $M$ the group  becomes $C_{2v}$ consisting of $\{E, U_2,\sigma_h, \sigma_v^\prime\}$.
    Along the lines $\Gamma-K$ and $M-K$ the group is $C_s$ consisting of $\{E,\sigma_h\}$ and
    along $\Gamma-M$ it is the same as at $M$.
    
    Finally, making the system both buckled and breaking the inversion.
    Then the group at $\Gamma$ is only $D_3$, containing $\{E,2C_3, 3U_2\}$.  At $K$ it becomes $C_3$ at $M$ and along $\Gamma-M$ it becomes $C_2$
    with only elements $\{E,U_2\}$ along $\Gamma-K$ and $M-K$ there is no symmetry left at all.
    
    The character table of $C_3$ is given in Table \ref{tabC3}
    \begin{table}
    	\caption{Character table of $C_3$, $\omega=e^{2i\pi/3}$. \label{tabC3}}
    	\begin{ruledtabular}
    		\begin{tabular}{ll|rrr}
    			$C_3$ & & $E$ & $C_3$ & $C_3^{-1}$ \\ \hline
    			$K_1$ & $A$ & 1 & 1 & 1 \\
    			$K_2$ & $E$ & 1 & $\omega$ & $\omega^*$ \\
    			$K_3$ & $E$ & 1 & $\omega^*$ & $\omega$ \\
    		\end{tabular}
    	\end{ruledtabular}
    \end{table}

    We note that in the group $C_3$, Koster \etal \cite{Koster} labels the two irreps which cannot be made real  as two separate irreps while
    in the 'chemical' notation, they are both labeled $E$. This is because if one ignores spin these two irreps are each other's complex conjugate
    and become degenerate by time reversal. They form a Kramers doublet.
    However, taking into account the spin $1/2$ no degeneracy between the two occurs because time reversal takes spin up in to spin down.
    Adding the horizontal mirror plane just
    adds another label $^\prime$ or $^\dprime$ for even or odd under that operation. Thus we can see that that the degenerate
    levels $E$ in the buckled case or $E^\prime$ and $E^\dprime$ would be allowed to split and open a gap.
    This is well known to open the gap in the honeycomb BN case. 
    
    \subsection{Tight-binding model} \label{sec:tb}
    
    \begin{figure}[!htb]
    	\includegraphics[width=8cm]{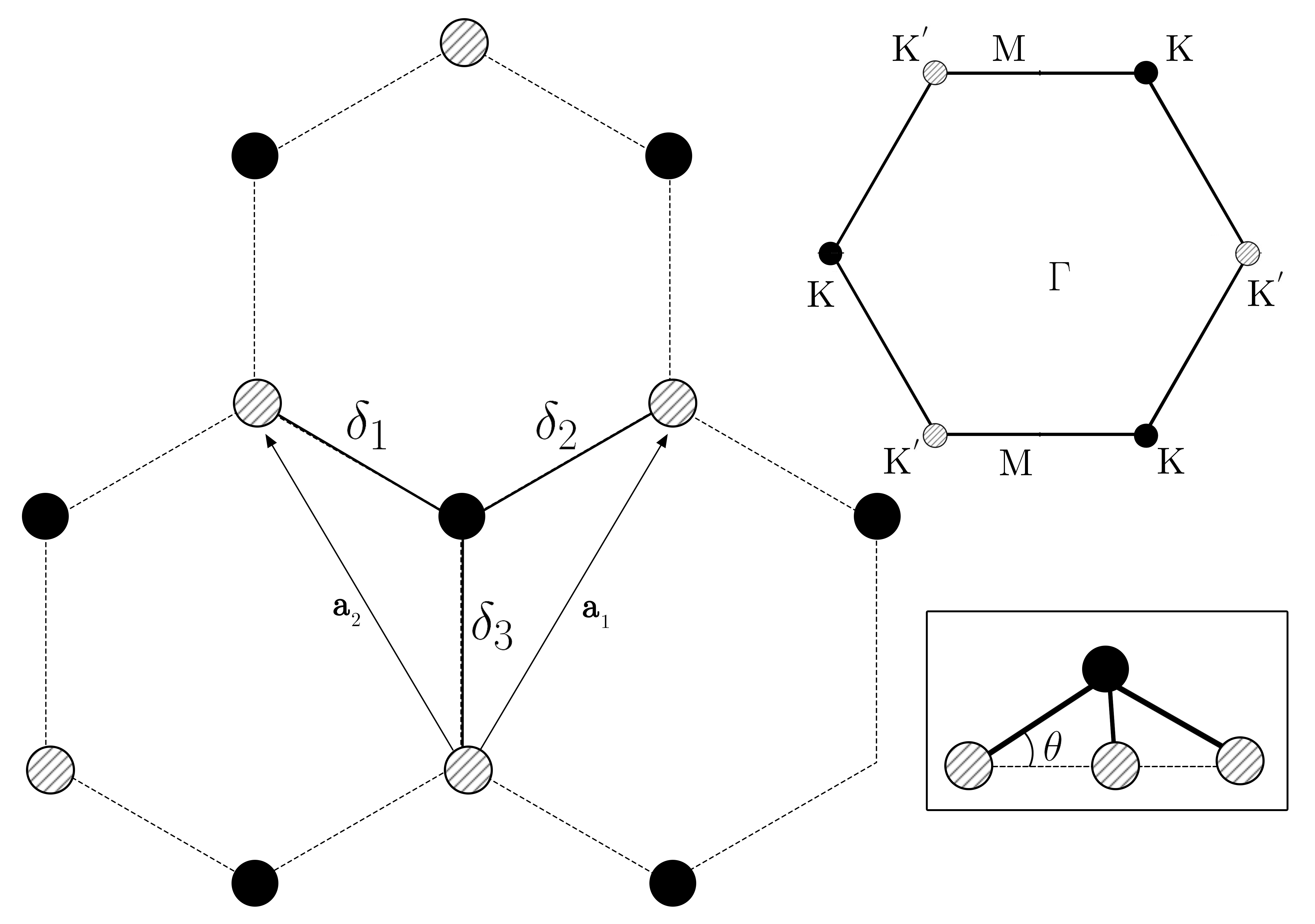}
    	\caption{Lattice of low buckled antimonene. Shaded and dark circles represent the Sb at different height. Inset shows the definition of buckling used in the paper along with the corresponding $1^{st}$ $BZ$.\label{fig:honeycomb}}
    	
    \end{figure}
    In this section, we construct a nearest neighbor tight-binding Hamiltonian
    for the Sb-$p$ derived orbitals. The structure is shown in Fig.
    \ref{fig:honeycomb}. 
    For the bands near the Fermi level, 
    one can ignore the contribution of $s$ states as they occur
    at much lower energy as seen in the previous section.
    Separating the orbitals according to their sublattice $A$, $B$,
    the Hamiltonian takes the block form: 
    \begin{align}
    H=\bordermatrix{~ &A & B\cr
    	A & \bar{\Delta}+\mu_A I_{3\times 3} & \bar{H} \cr
    	B & \bar{H}^* & \bar{\Delta}+\mu_B I_{3\times 3} \cr}\label{eq:1}
    \end{align}
    Here
    \begin{equation}
    \bar{\Delta}=\left[ \begin{array}{ccc} \Delta_\pi & & \\
    & \Delta_\pi & \\ & & \Delta_z 
    \end{array}\right]
    \end{equation}
    gives the energy shift of the atomic $p$-orbitals from the zero reference
    energy. We included here the fact that the on-site diagonal energy
    for $p_z$ orbitals may be different from that of $\{p_x,p_y\}$ orbitals.
    We can also switch on a different potential ($\mu_A,\mu_B$) 
    on each sublattice and examine which band crossings open up as a gap
    in response. The off-diagonal $AB$-blocks involve the Bloch sum over the
    three nearest neighbors:
    \begin{gather}
    \begin{split}
    \bar{H}&=T_1(V_{\pi},V_{\sigma},\theta)e^{i{\bf k}\cdot\bm{\delta}_1}+T_2(V_{\pi},V_{\sigma},\theta)e^{i{\bf k}\cdot\bm{\delta}_2}\\
    &+T_3(V_{\pi},V_{\sigma},\theta)e^{i {\bf k}\cdot \bm{\delta}_3} \label{eq:3}
    \end{split}
    \end{gather}
    where
    \begin{eqnarray}
    \bm{\delta}_1&=&(-\cos{\theta}\sqrt{3}/2,\cos{\theta}/2,\sin{\theta})a/\sqrt{3}, \nonumber \\ 
    \bm{\delta}_2&=&(\cos{\theta}\sqrt{3}/2,\cos{\theta}/2,\sin{\theta})a/\sqrt{3}, \nonumber \\
    \bm{\delta}_3&=&(0,-\cos{\theta},\sin{\theta})a/\sqrt{3}
    \end{eqnarray}
    in terms of the lattice constant $a$ 
    as shown in \autoref{fig:honeycomb}.
    The side of the hexagon is $a/\sqrt{3}$.
    Here $T_1$, $T_2$, $T_3$
    can be expressed in terms of the $V_\pi$ and $V_\sigma$ components of the nearest neighbor hopping interaction using
    the Koster-Slater two-center approximation.\cite{SlaterKoster} We assume here  that under buckling, the bond distance $a$ is kept fixed
    and the buckling angle just changes the relative contributions of $V_\pi$ and $V_\sigma$ to the hopping integrals.
    We chose the $V_\pi/V_\sigma$ ratio to be $-0.275$. This value
    is close to Walter Harrison's universal ratio.\cite{Froyen} Together with the
    $\Delta_\pi-\Delta_\sigma$ parameters, they were chosen to
    present a similar band structure to the DFT results of the main paper.
    However, a detailed fit was not attempted because the actual
    band structure is affected by interaction with nearby $d$ and $s$ bands
    and the main purpose of our tight-binding model is to study
    the generic behavior of the bands as function of buckling angle and
    mostly the topological features.
    
    Each $T_i$ is a $3\times3$ matrix in the basis of the $p_m$, $m=(x,y,z)$ orbitals.
    
    \begin{widetext}
    	\begin{equation}
    	T_1(V_{\pi},V_{\sigma},\theta)=\left[\begin{matrix}V_{\pi} \left(- \frac{3 \cos^{2}{\theta}}{4} + 1\right) + V_\sigma \frac{3  \cos^{2}{\theta}}{4} &  \left( V_{\pi} - V_{\sigma}\right) \frac{\sqrt{3} \cos^{2}{\theta}}{4} &  \left(V_{\pi} - V_{\sigma}\right)\frac{\sqrt{3} \sin{\theta} \cos{\theta}}{2}\\ \left(V_{\pi} - V_{\sigma}\right)\frac{\sqrt{3} \cos^{2}{\theta}}{4} & V_{\pi} \left(- \frac{\cos^{2}{\theta}}{4} + 1\right) + V_\sigma\frac{\cos^{2}{\theta}}{4} & \left( V_{\sigma} - V_{\pi}\right)\frac{\sin{\theta} \cos{\theta}}{2}\\ \left(V_{\pi} - V_{\sigma}\right)\frac{\sqrt{3}  \sin{\theta} \cos{\theta}}{2} & \left( V_{\sigma} - V_{\pi}\right)\frac{ \sin{\theta} \cos{\theta}}{2} & V_{\pi} \cos^{2}{\theta }  + V_{\sigma} \sin^{2}{\theta}\end{matrix}\right] 
    	\end{equation}
    	
    	\begin{equation}
    	T_2(V_{\pi},V_{\sigma},\theta)=\left[\begin{matrix}V_{\pi} \left(- \frac{3 \cos^{2}{\theta}}{4} + 1\right) + V_{\sigma} \frac{3 \cos^{2}{\theta}}{4} & \left( V_{\sigma}- V_{\pi} \right)\frac{\sqrt{3}  \cos^{2}{\theta}}{4} &  \left( V_{\sigma}- V_{\pi} \right)\frac{\sqrt{3} \sin{\theta} \cos{\theta}}{2}\\  \left( V_{\sigma}- V_{\pi} \right)\frac{\sqrt{3} \cos^{2}{\theta}}{4} & V_{\pi} \left(- \frac{\cos^{2}{\theta}}{4} + 1\right) + V_{\sigma}\frac{ \cos^{2}{\theta}}{4} & \left( V_{\sigma} - V_{\pi} \right)\frac{ \sin{\theta} \cos{\theta}}{2}\\ \left( V_{\sigma} - V_{\pi}\right)\frac{\sqrt{3} \sin{\theta} \cos{\theta}}{2} & \left( V_{\sigma}- V_{\pi}\right)\frac{ \sin{\theta} \cos{\theta}}{2} & V_{\pi} \cos^{2}{\theta}  + V_{\sigma} \sin^{2}{\theta}\end{matrix}\right]
    	\end{equation}
    	
    	\begin{equation}
    	T_3(V_{\pi},V_{\sigma},\theta)=\left[\begin{matrix}V_{\pi} & 0 & 0\\0 & V_{\pi} \sin^{2}{\theta} + V_{\sigma} \cos^{2}{\theta} & \left(V_{\pi} - V_{\sigma}\right) \sin{\theta} \cos{\theta}\\0 & \left(V_{\pi} - V_{\sigma}\right) \sin{\theta} \cos{\theta} & V_{\pi} \cos^{2}{\theta} + V_{\sigma} \sin^{2}{\theta}\end{matrix}\right]
    	\end{equation}
    	
    \end{widetext}

    It is insightful to first understand the unbuckled case ($\theta=0$). In that case, the $p_z$ orbitals are decoupled from the
    $p_x,p_y$ orbitals because the former are odd with respect to the horizontal mirror plane and the latter are even.
    
    The $z$-part of the Hamiltonian becomes the well-known
    \begin{equation}
    H_z({\bf k})=\left[ \begin{array}{cc}\Delta_z & V_\pi g_0({\bf k})\\  V_\pi g^*_0({\bf k}) & \Delta_z \\
    \end{array}\right]
    \end{equation}
    with
    \begin{equation}
    g_0({\bf k})=\sum_{j=1}^3 e^{i{\bf k}\cdot{\bm{\delta}_j}}
    \end{equation}
    giving the purely $\pi$-bonded bands.
    At $\Gamma$ we find $E({\bf k}_\Gamma)=\Delta_z \pm 3V_\pi$ with eigenvectors
    $\frac{p_{zA}\mp p_{zB}}{\sqrt{2}}$. At $K$, we have
    $e^{i{\bf k}\cdot\bm{\delta}_1}=e^{-i2\pi/3}$, $e^{i{\bf k}\cdot\bm{\delta}_2}=e^{i2\pi/3}$
    $e^{i{\bf k}\cdot\bm{\delta}_1}=1$ and $g_0({\bf k})=0$ giving the
    doubly degenerate eigenvalue $\Delta_z$. This is the Dirac point $E^\dprime$.

    The $xy$ part of the Hamiltonian has the off-diagonal part 
    \begin{equation}
    \bar{H}_{xy}=\left [ \begin{array}{cc}\frac{3V_\sigma+V_\pi}{4}g_+({\bf k})
    +V_\pi e^{i{\bf k}\cdot\bm{\delta}_3} &
    \frac{\sqrt{3}(V_\pi-V_\sigma)}{4} g_-({\bf k})   \\
    \frac{\sqrt{3} (V_\pi-V_\sigma)}{4} g_-({\bf k}) &
    \frac{3V_\pi+V_\sigma}{4}g_+({\bf k})
    +V_\sigma e^{i{\bf k}\cdot\bm{\delta}_3}
    \end{array}\right ]
    \end{equation}
    with $g_\pm({\bf k})=e^{i{\bf k}\cdot\bm{\delta}_1}\pm e^{i{\bf k}\cdot\bm{\delta}_2}$.
    At $\Gamma$, the $g_-({\bf k_\Gamma})=0$ and we obtain two degenerate eigenvalues
    $\Delta_\pi\pm (V_\sigma+V_\pi)3/2$ with eigenstates $(p_{xA}\mp p_{xB})/\sqrt{2}$
    and $(p_{yA}\mp p_{yB})/\sqrt{2}$.  At $K$, the matrix can still be
    diagonalized analytically.
    The off-diagonal part here takes the form
    \begin{equation}
    \bar{H}_{xy}({\bf k}_K)=\left [ \begin{array}{cc}
    -\frac{3}{4}(V_\sigma-V_\pi)& i\frac{3}{4}(V_\sigma-V_\pi) \\
    i\frac{3}{4}(V_\sigma-V_\pi) & \frac{3}{4}(V_\sigma-V_\pi) 
    \end{array}\right]
    \end{equation}
    The Hamiltonian then has a double degenerate eigenvalue $E=\Delta_\pi$ with
    eigenvectors $\pi_A=(p_{xA}+ip_{yA})/\sqrt{2}$ and
    $\pi_B^*=(p_{xB}-ip_{yB})/\sqrt{2}$, the Dirac point $E^\prime$, 
    and two non-degenerate eigenvalues
    $\Delta_\pi+(V_\pi-3V_\sigma)/2$ with eigenvector $(\pi_A^*-\pi_B)/\sqrt{2}$
    $\Delta_\pi-(V_\pi-3V_\sigma)/2$) with eigenvector $(\pi_A^*+\pi_B)/\sqrt{2}$.
    In other words, it can be diagonalized in the basis of the
    $\pi_A$, $\pi_A^*$, $\pi_B$ and $\pi_B^*$ orbitals.\cite{wu_sarma_2008}
    
    We now examine the band surfaces in 3D, in particular the
    intersection of the down pointing Dirac cone derived from
    the $p_\pi$ orbitals and the upward pointing Dirac cone derived
    from the $p_z$ orbitals. 
    To further study this crossing  analytically  we expand
    the tight-binding Hamiltonian around $K$, \ie for ${\bf k}={\bf k}_K+{\bf q}$
    for small ${\bf q}=(q\cos{\phi},q\sin{\phi})$. The azimuthal angle $\phi$ of the ${\bf q}$
    is measured from the $X$-direction for the $K$-point along $x$
    and $q=|{\bf q}|$. 
    
    The eigenvalues are symmetric about the $\Delta_z$ and $\Delta_\pi$
    and given by
    \begin{eqnarray}
    E_z&=&\Delta_z\pm v_z q \pm \frac{q^2}{m_z}\cos{(3\phi)}, \nonumber \\
    E_\pi&=&\Delta_\pi\pm v_\pi q \pm \frac{q^2}{m_\pi}\cos{(3\phi)} \label{eqcone}
    \end{eqnarray}
    Here $v_z$, $v_\pi$ are the Dirac linear dispersion velocities a
    and $m_z$, $m_\pi$ are an effective mass parameter. The actual effective masses
    depend on the direction of ${\bf q}$ leading to a warping of the
    constant energy lines with three-fold symmetry, while the velocity is
    isotropic. 
    For the $p_z$ Hamiltonian, one finds in our
    nearest neighbor tight binding Hamiltonian, $v_z=\sqrt{3}\pi V_\pi$
    and $m_z^{-1}=-V_\pi \pi^2/2$. Thus both are completely determined by
    the $V_\pi$-interaction between $p_z$ orbitals. On the other
    hand, one may also keep $m_z$ and $v_z$ as independent parameters
    to make the effective low energy Hamiltonian and eigenvalues
    applicable beyond the tight-binding model. Their form is
    dictated by symmetry.
    
    \begin{figure}[!htb]
    	\includegraphics[width=8cm]{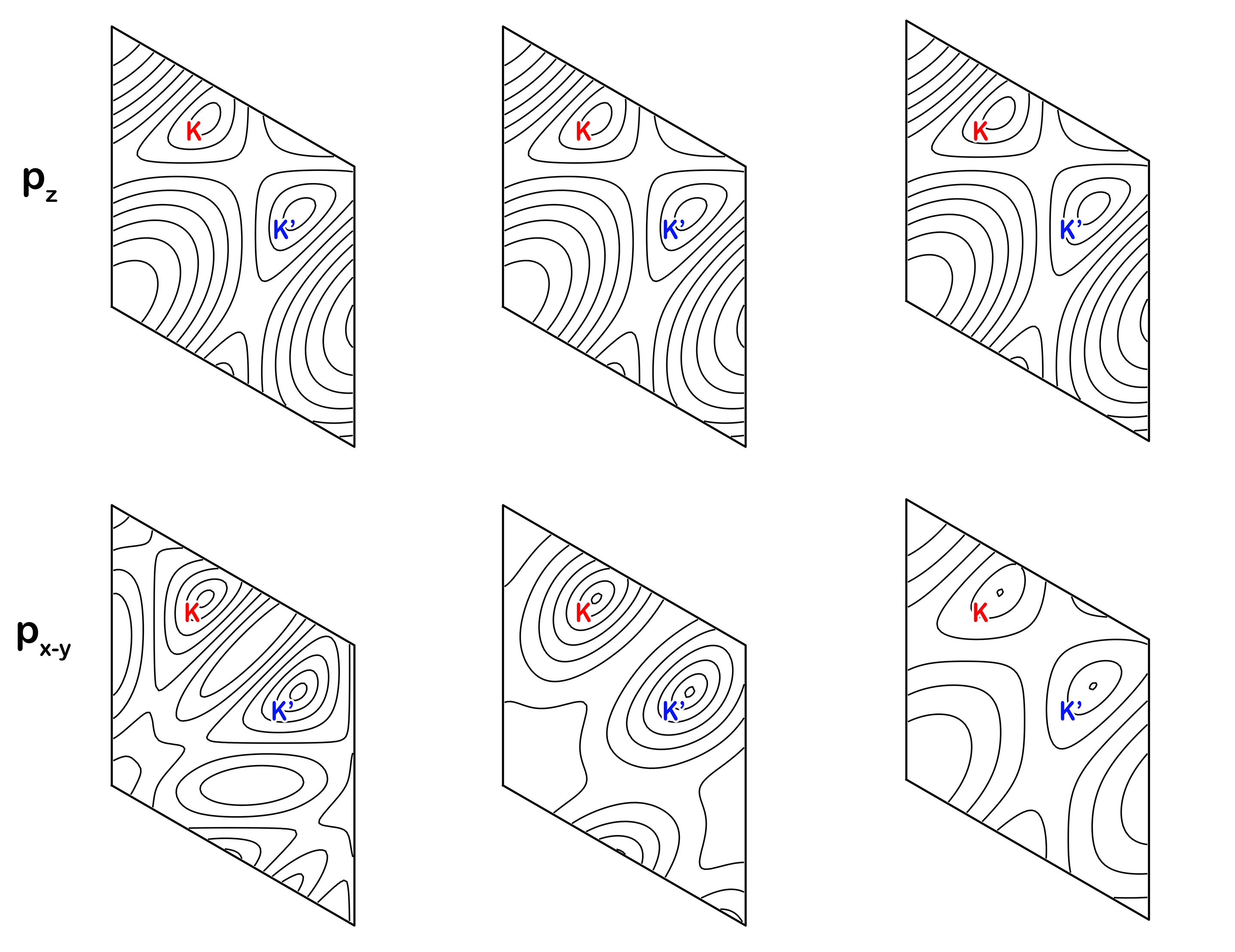}
    	\caption{Contour plot of valence  $p_z$ (top) and 
    		conduction $\{p_x,p_y\}$   (bottom) bands
    		for various values of $(V_\pi,V_{\sigma})$. From left to right,
    		$(V_\pi,V_{\sigma})$ =($-0.66$,2.4),($-0.66$,3.6),($-0.66$,4.44) eV. 
    		$K$ and $K^\prime$ points are marked. The $\Gamma$-point occurs at each of the corners of the 
    		reciprocal unit cell spanned by ${\bf b}_1$,${\bf b}_2$. 
    		\label{fig:vpi-vsig}}
    \end{figure}
    
    For the $\{p_x,p_y\}$ derived bands, the eigenvalues of the tight-binding
    Hamiltonian are also
    found to be symmetric about $\Delta_\pi$  and are found by
    diagonalizing $H_{eff}=\bar{H}_{xy}\bar{H_{xy}}^\dagger$ and taking
    $\pm\sqrt{\lambda}$ of its eigenvalues $\lambda$. These eigenvalues
    were worked out by Wu  \etal\cite{wu_sarma_2008} for the special
    case that $V_\pi=0$, which applies to optical lattices. In that
    case, two bands are found to be flat, and two are Dirac cone like
    and have exactly the same shape  as for the $p_z$ orbitals.
    With a mixture of both $V_\sigma$ and $V_\pi$, the expressions
    of the expansion in $q$ 
    become too complex 
    to be useful but the important finding is that by choosing 
    the $m_\pi$ of opposite sign as the $m_z$ orbitals  the warping
    is found to be rotated by 30$^\circ$.
    In the DFT band structure in the main paper
    we can see that the particle-hole symmetry about $\Delta_\pi$
    which is the $E^\prime$  value above the Fermi level, no longer holds.
    This is because of the interactions with the higher lying bands which
    are derived from the Sb-$d$ orbitals and not included in the model.
    Therefore, it  is not that useful to find expressions for the
    velocity and mass parameter of this Dirac cone in terms of the
    tight-binding model because the latter has only limited validity.
    What matters is that both Dirac cone bands near $K$ and
    extending up to the region of their intersection
    can be described by Eq. \ref{eqcone} which represents two trigonally
    warped Dirac cones and that in the DFT results these cones
    are found to be rotated 30$^\circ$ with respect to each other. 
    
    \begin{figure*}[!htb]
    	\includegraphics[width=18cm]{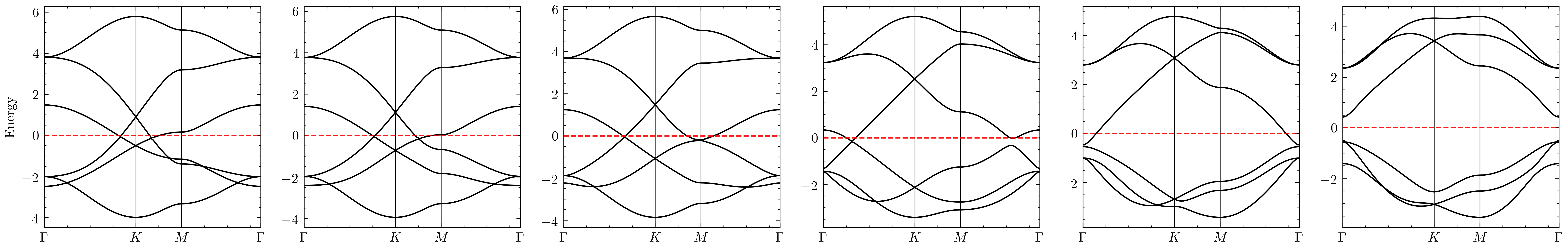}
    	\caption{Evolution of band structure at different buckling
    		angles $\theta$. From left to right) $0^\circ$, $5^\circ$, $7.34^\circ$, $20^\circ$, $36^\circ$. In this figure the energy units are eV
    		and $V_\sigma=2.4$ eV, $V_\pi=-0.66$ eV. The red-dashed line
    		at 0 energy is the Fermi energy in the flat case but is kept
    		fixed in the later figures without adjusting the Fermi energy.
    		In the before last panel, the Fermi energy lies exactly at the degenerate band touching at $\Gamma$. 
    		\label{fig:band_evolution}}	
    \end{figure*}
    
    Examining the bands in Fig. 1 in the main paper one can see
    that the upper band of the $p_z$ derived Dirac cone deviates upward
    from the linear behavior and thus has positive $m_z$ which agrees
    with the $V_\pi<0$. The lower band of the $p_\pi$ derived Dirac cone
    can be seen to bend down and thus has the opposite mass $m_\pi<0$.
    Note that the direction $K-\Gamma$ corresponds to $\phi=\pi$.
    and the direction $K-M$  corresponds to $\phi=2\pi/3$.

    Both of the Dirac cones can be obtained from an effective low Hamiltonian
    of the form
    \begin{equation}
    H_{eff}=\left[ \begin{array}{cc} \Delta_z & qv_z+q^2\frac{\cos{(3\phi)}}{m_z} \\ qv_z+q^2\frac{\cos{(3\phi)}}{m_z} & \Delta_z
    \end{array}\right ] \label{eq:heff}
    \end{equation}
    with a similar one for the $p_\pi$ case.

    To illustrate the behavior of the Dirac cone warping, we show in
    Fig. \ref{fig:vpi-vsig}  contour plots obtained in the tight-binding Hamiltonian
    for different choices of $V_\sigma,V_\pi$ for the $p_z$ derived
    and $\{p_x,p_y\}$ derived cones. One can see that while for the $p_z$ derived
    ones the corners of the triangular contours around $K$ are point in
    the $K-M-K^\prime$ direction and the shape does not depend on $V_\sigma$
    because these bands only involve $V_\pi$.  In contrast the $\{p_x,p_y\}$
    derived cones for the first choice of $V_\sigma,V_\pi$ parameters which
    best matches the DFT bands and are used in the main paper, the cones
    are rotated by 30$^\circ$ with respect to the $p_z$-cone. The flat edge
    of the triangle is now along the $K-M-K^\prime$ direction. However,
    as we change the $V_\sigma,V_\pi$ these cones can become almost circular
    (middle case) or their warping rotated the same as for $p_z$  as $V_\sigma$
    is increased. Thus the rotated warping of the $p_z$ relative to
    the $\{p_x,p_y\}$ derived cones is sensitive to the relative values of
    $V_\sigma$ and $V_\pi$ and this is what is ultimately responsible for
    the shape of the nodal line
    and the occurrence of six Dirac points after buckling.


    The effects of buckling in our model are incorporated through the
    $\theta$ dependence. As already explained in the main text, when
    $\theta\ne0$ the off-diagonal terms between the $p_z$ block and $\{p_x,p_y\}$
    blocks are turned on. The resulting $6\times6$ matrix is readily
    diagonalized numerically but analytic expressions are no longer
    possible. Instead in the main text we then focus on the effective low
    energy Hamiltonian near the nodal line, which explains its breaking up
    into six separate Dirac points. 
    
    In Fig. \ref{fig:band_evolution} we show the tight-binding
    bands of the $6\times6$ $p$-orbital derived Hamiltonian for various
    buckling angles. The leftmost figure for the flat case can be compared with
    the DFT results given in Fig. 1(a) of the main paper. 
    The $V_\sigma$, $V_\pi$ and $\Delta_\pi$, $\Delta_z$
    parameters were chosen to give about the correct ratios of the
    splitting of the two Dirac points $E^\prime$, $E^\dprime$ at $K$,
    the splitting of the outermost
    eigenvalues $A_2^\prime$, $A_1^\prime$ (see labeling of Fig. 1(a) in main paper)
    of the $\{p_x,p_y\}$ derived bands at $K$ and the splitting of the $p_z$ derived
    bands at $\Gamma$. Note that both the $\{p_x,p_y\}$ derived bands
    and $p_z$ derived bands are `particle-hole' symmetric about their
    center of gravity, the $E^\prime$ and $E^\dprime$ Dirac points at $K$,
    which are displaced from each other by $\Delta_\pi-\Delta_z$ for
    the flat case.  We can see
    that the lower band levels  $A_{2u}$, $E_{2g}$ at $\Gamma$ are then close to each other  while the upper band levels at $\Gamma$, $B_{2g}$ and $E_{1u}$
    are more separated and inverted from the
    DFT bands. 
    We are not trying to reproduce the DFT bands exactly because
    these upper bands are influenced by the interaction with the Sb-$d$ bands
    in the DFT results.  Our main goal here is just to see the
    qualitative evolution of the bands under buckling. 
    
    We can see the
    Dirac points move toward $M$, disappear at $M$ and then the remaining ones
    move closer to $\Gamma$ and finally the full gap opens.

    \subsection{Winding number}\label{sec:winding}
    One can calculate the winding number of the Dirac cones around a contour $C$
    around the  Dirac point given by \cite{Goerbig2017}
    \begin{align}
    W_c=\frac{1}{2\pi}\oint_C \nabla_q\phi_{\vec{q}} \cdot dq
    \label{eq:7}
    \end{align}
    where $\phi_{\vec{q}}$ is the relative phase between the two
    coefficients in the eigenvector $\tan(\phi_{\vec{q}}) =\frac{u_2}{u_1}$
    of the  $2\times2$ effective Hamiltonian near the Dirac point (Eq.(2) in the main paper). Alternatively, in the tight-binding model, we can
    open up a gap at each of these Dirac cones by adding the $\mu_A$, $\mu_B$
    parameters giving a different on-site energy to the $A$ and $B$ atoms in the unit cell.  Once a small gap is opened up, the Berry curvature, \ie the curl of
    the Berry connection, is calculated numerically on a  fine {\bf k}-mesh and 
    for each occupied state. Summing these gives the total winding number
    as shown in Fig. 2(b) in the main paper. 
    Finally, also in the tight-binding model, we can calculate the accumulated
    Berry phase along a small contour around the Dirac point. 
    We have verified that these different procedures agree with each other.
    
    \subsection{Results for arsenene}\label{sec:As}
    \begin{figure}[!htb]
    	\includegraphics[width=5 cm]{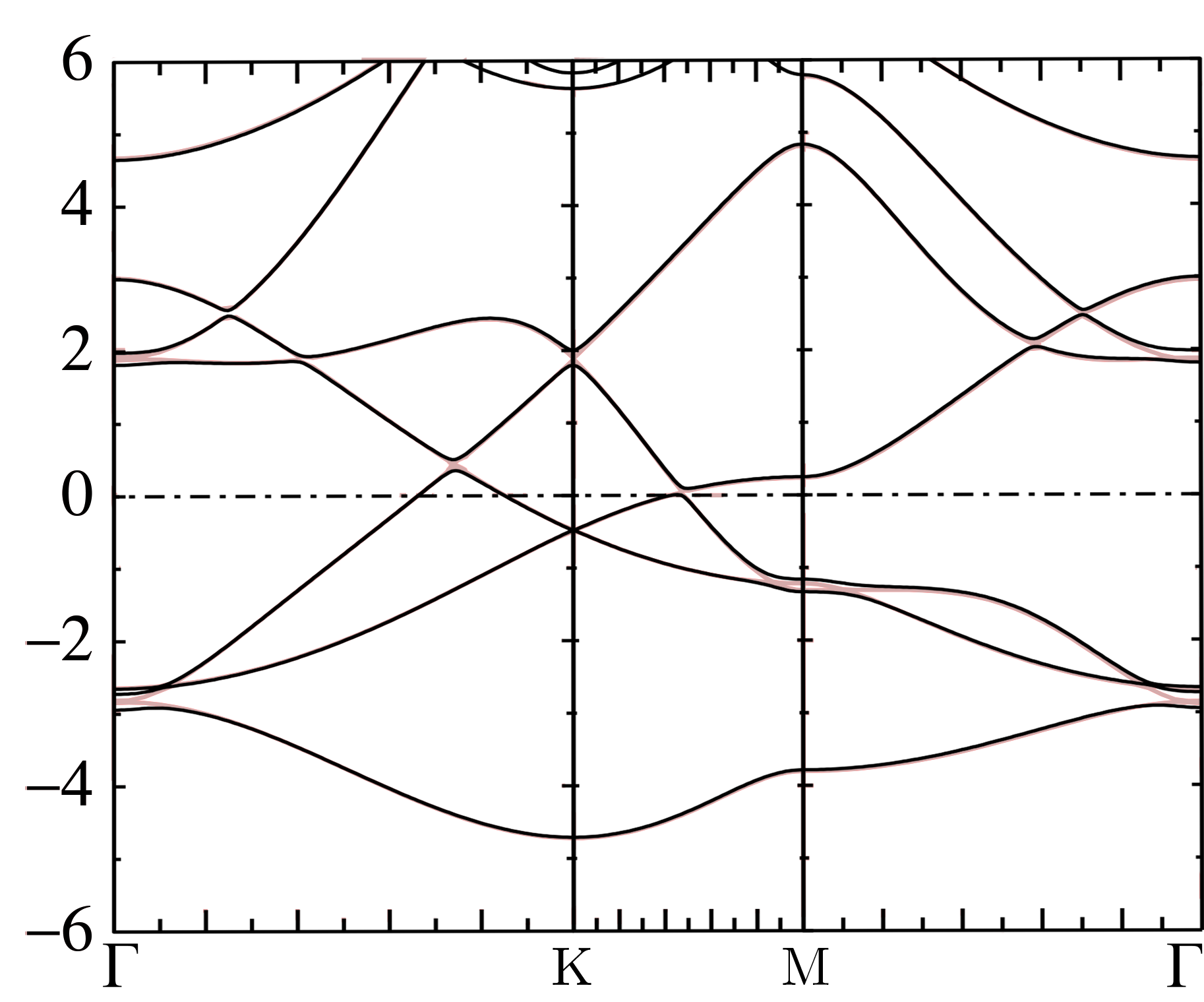}
    	\caption{Band structure of arsenene (As)
    		with (black) and without (red) SOC at the LDA level.
    		\label{fig:As-soc}}	
    \end{figure}
    
    While the main paper is mostly focused on antimonene (Sb), very similar
    results hold for arsenene (As). The relative stability of the flat
    and buckled honeycomb structure was already shown in Fig. \ref{fig:tote}.
    The band structure at the LDA level is shown in Fig. \ref{fig:As-soc}
    both with and without spin-orbit coupling.  We can see that the Fermi level
    again occurs near the crossing of the two Dirac cones centered at $K$.
    The crossings of these cones along $\Gamma-K$ and $K-M$
    are again tilted with respect to each other. The spin-orbit coupling
    opens a gap at all the Dirac points. We can see that as in Sb,
    the spin-orbit opened gap is larger for the upper $E^\prime$ $\{p_x,p_y\}$
    derived Dirac point at $K$ than for the lower
    $E^\dprime$ $p_z$ derived Dirac point. At the new Dirac points along
    $\Gamma-K$ and $K-M$, the splitting is intermediate. The main point is
    that the gaps opened here at the Dirac points near $E_F$
    are straddled with respect to each other so that the system remains
    overall metallic, in contrast with the Sb case in the main paper Fig. 3,
    where a zero indirect gap situation emerges.

\end{document}